\documentclass[11pt, a4paper,hidelinks]{article}
\usepackage[margin=1in]{geometry}
\usepackage{epsfig,amssymb,soul}
\usepackage{hyperref}
\usepackage{graphicx}
\usepackage{epstopdf}
\usepackage{amsfonts,amsmath}
\usepackage{bm}
\usepackage{setspace}
\usepackage{comment}
\usepackage{siunitx}
\usepackage{mathrsfs}
\usepackage{multirow}
\usepackage[hang,flushmargin,bottom]{footmisc}
\usepackage[normalem]{ulem}
\usepackage{mathtools}
\usepackage{lipsum}
\usepackage{array}
\usepackage{enumerate}
\usepackage{ifthen}
\usepackage[affil-it]{authblk}
\usepackage{cite}
\usepackage{leftidx}
\usepackage{relsize}
\usepackage{braket}
\usepackage{float}

\usepackage[autostyle]{csquotes}
\usepackage[dvipsnames]{xcolor}
\usepackage{comment}
\usepackage{color}
\usepackage{subcaption}
\usepackage{blkarray, bigstrut}

\newcommand\blfootnote[1]{%
	\begingroup
	\renewcommand\thefootnote{}\footnote{#1}%
	\addtocounter{footnote}{-1}%
	\endgroup
}

\providecommand{\keywords}[1]
{
	\small	
	\textbf{\textit{Keywords---}} #1
}

\title{Finite Element Network Analysis: \\ A Machine Learning based Computational Framework for the Simulation of Physical Systems}
\author[a]{Mehdi Jokar\thanks{email: mjokar@purdue.edu}}
\author[a]{Fabio Semperlotti}
\affil[a]{School of Mechanical Engineering, Ray W. Herrick Laboratories, Purdue University, West Lafayette, IN 47907}

\begin{document}

\maketitle
\begin{abstract}
	This paper introduces the concept of finite element network analysis (FENA) which is a physics-informed, machine-learning-based, computational framework for the simulation of physical systems. The framework leverages the extreme computational speed of trained neural networks and the unique transfer knowledge property of bidirectional recurrent neural networks (BRNN) to provide a uniquely powerful and flexible computing platform. One of the most remarkable properties of this framework consists in its ability to simulate the response of physical systems, made of multiple interconnected components, by combining individually pre-trained network models that do not require any further training following the assembly phase. This remarkable result is achieved via the use of key concepts such as transfer knowledge and network concatenation. Although the computational framework is illustrated and numerically validated for the case of a 1D elastic bar under static loading, the conceptual structure of the framework is extremely general and it suggests potential extensions to a broad spectrum of applications in computational science. 
	The framework is numerically validated against the solution provided by traditional finite element analysis and the results highlight the outstanding performance of this new concept of computational platform.
\end{abstract}

\section{Introduction}
Numerical simulations\blfootnote{accepted for publication in Computers and Structures} of physical systems play a vital role in computational physics and engineering. Over the past many decades, a plethora of numerical methodologies have been developed in support of the most diverse fields of science.
Arguably one of the most popular and versatile approach to the numerical simulations of linear and nonlinear systems has been the finite element (FE) method. The literature on this topic is extensive but excellent summaries of fundamental and seminal works can be found in~\cite{babuska2010finite,hughes2012finite,reddy2014introduction,zienkiewicz2005finite}. From a general perspective, the FE method owes its popularity to the great versatility and ability to model physical systems via a straight forward decomposition in elements of the entire computational domain. This approach allows a very natural treatment of multiple dissimilar elements, complex geometries, and diverse boundary conditions. 
Although the FE method has proven to be a powerful and reliable method to model a variety of physical phenomena mathematically described by integer-order differential equations, its computational cost increases very rapidly with the number of degrees of freedom (DOF)~\cite{Babuska1997-CvsDoF}. In fact, while FE models are proven to converge to the exact solution during a mesh refinement process, the price to pay is a rapid increase in resources and computational time due to increased number of DOFs. This fundamental characteristic of FE analysis is also at the basis of some important limitations, particularly in the area of large scale models and multiscale analysis~\cite{argilaga2018fem-MultiScale}. A few relevant examples include multiscale mechanics (such as damage initiation and propagation~\cite{rafii1998multi, sih2001mesofracture}, nonlocal and scale effects~\cite{giacomini2005ambrosio, abraham1998spanning}, turbulent flows~\cite{rasthofer2018-MultiscaleTurbulent}), high-frequency wave propagation~\cite{engquist2003computational} , and contact dynamics~\cite{style2018contact}.

The need for better performance in these demanding areas has led the scientific community to explore modified versions of finite element analysis (FEA) that could address more directly and efficiently these important shortcomings. Formulations like the spectral finite elements~\cite{SFEM}, hp-FEM~\cite{hpFEM}, model reduction and super-elements~\cite{FEM_order_reduc}, and extended finite element method~\cite{XFEM} have specifically targeted both large scale and multiscale computations.
Although many of these methods have been extremely successful in significantly expanding the capabilities of FEA, and despite the outstanding advances in computational resources and methods (e.g. parallel or distributed memory), there are several classes of problems that are still out of reach for FEA. A few examples include multiscale problems spanning a wide range of spatial (e.g. porous materials, biomedical tissues and bones, damage initiation and evolution) and temporal scales (e.g. high frequency wave propagation, impacts, contacts)~\cite{matouvs2017review, tong2019review}. 
We merely note that, although the above discussion was cast in a perspective of FE methods, most of the mentioned limitations readily extend to other popular numerical techniques such as boundary element methods, finite difference method, and spectral methods~\cite{han2006spline, guermond2009nonlinear, SFEM}. 

It appears that a paradigm shift is needed in order to develop computational methods that, while retaining the many advantages of FEA (particularly, the ability to solve systems of differential equations over domains having complex geometries and boundary conditions), are capable of extreme computational efficiency over large (potentially multiscale) domains. In other terms, a new class of computational methods should allow extreme scalability with minimum impact on the computational time, a property that we refer to as \textit{embarrassingly scalable} models.
Most of these computational challenges of the methods previously described are rooted in the same practical issue: the rapid increase in the size of the system of governing equations with the increasing number of degrees of freedom. It follows that a possible route towards a powerful computational framework could be to bypass entirely the direct numerical solution of the governing equations. Analytical solutions could certainly be very powerful alternative but they are typically not achievable on complex shaped domains and boundary conditions. Therefore, the fundamental question of how to efficiently solve these systems of equations remains open.

In recent years, the advent of machine learning algorithms and the ongoing revolution in data-driven techniques has offered an outstanding opportunity to pursue novel directions for numerical simulations~\cite{NNReview,NN-Nature1,NN-Nature2,NN-3}. Several recent studies have shown that trained neural network (NN) models can efficiently predict the response of large, complex, and even nonlinear systems with unprecedented computational efficiency~\cite{NNReview, gu2018recent-CNN}. This result is enabled by the intrinsic nature of a neural network which, once trained to mimic the response of a physical system, does not require to solve an actual system of differential equations in order to evaluate the response of the system to a new set of input.
To-date, many different architectures of neural networks have been proposed to solve a wide range of problems~\cite{NN-Deep_Review,NN-DeepBook}. 
Most data-driven training methods for NN require extensive training data sets and can only simulate very specific conditions~\cite{liang2018deep}.
In order to compensate for the limited flexibility of trained networks in modeling different systems, some researchers explored the possibility to combine the concepts of FEA as a general solver and of a neural network as a computationally efficient framework. This approach gave rise to the idea of finite element neural networks (FENN)~\cite{FENN1,FENN2,FENN4}. Takeuchi \textit{et al.}~\cite{FENN4} embedded the FE conceptual approach into the network formulation and described how to construct the network architecture based on different finite elements and shape functions. 
Along a similar direction, Xu \textit{et al.}~\cite{FENN3} used a deep network to solve Maxwell's equations and simulate the magnetic flux leakage, while Liu \textit{et al.}~\cite{liu2019deep} modeled a heterogeneous material response to external load using an assembly of multiple NNs. In all these approaches, the network architecture
was pre-determined and constrained by the element types. Hence, every change made to the structural model (e.g. mesh size, boundary conditions, and material properties) required updating the entire architecture and re-training the network. This is a critical aspect because it is well-known that the training phase is the most time demanding operation in NN applications.
We also mention that, in FENN, the total number of neurons scales with $N^2$, where $N$ is the total number of FE nodes~\cite{FENN3}. It follows that this approach becomes quickly unmanageable for large scale computations because it gives rise to NN with an extremely large number of neurons.  
From a general perspective, trained neural networks are very computationally efficient when used for the prediction of the response of physical systems. This is due to the fact that a trained NN effectively provides an equivalent transfer function of the system (even in presence of complex interconnected and nonlinear systems) which does not require any further solution of the governing equations.
Despite this undeniable computational advantage, the main barrier limiting the large scale deployment of NN for computations lies in the need to redesign and retrain the network every time the physical parameters of the systems are changed. It is well-known that the training phase is particularly cumbersome and requires large data sets in order to obtain networks that are well representative of the physical systems.
Evidently, this latter aspect is very detrimental in a perspective of developing a generic NN-based computational framework for the simulation of physical systems.

For NN-based computations to become a viable alternative to FE techniques, the modeling strategy should be sufficiently flexible to allow interconnecting pre-trained NN models (representative of selected physical behaviors or systems) in order to build complex assemblies without any further training. In other terms, one could envision a general modeling approach where pre-trained NN models capable of simulating the physical behavior of selected structural elements (e.g. beams, plates, and shells) are available in an existing database. When assembling a model of a physical system, individual pre-trained NN modules can be selected from the database and interconnected together in order to create the final system assembly. This strategy is, of course, reminiscent of the general modeling approach in classical FEA where elements (described mathematically by shape functions, kinematic, and constitutive relations) are selected from a library and used to assemble complex models.

However, when using NN as fundamental building blocks of a physical system, the main obstacle consists in the need for a dedicated training phase for any given system configuration. Recall that NNs are typically trained under very strict conditions including, but not limited to, a specific configuration of the physical system and a fixed number of input variables. It follows that trained NNs typically do not allow for changes to the number and type of parameters that had not been previously defined as input. More recently, this problem was addressed by a new class of networks, known as recurrent neural networks (RNN)~\cite{Goodfellow-et-al-2016}, capable of learning the recursive logic of a given output parameter sequence with respect to independent sequential input. A practical example can be the ability to account for the temporal or spatial dependencies in the physical response of a system. 
This feature of RNNs is particularly suitable to develop network models that can handle a variable input size. In this way, the input can be adjusted based on the specific requirements of the system to be simulated including variable numbers of external loads, boundary conditions, and even constitutive components.

While the basic RNN relaxes the constraint on the fixed number of input, the network still operates according to a preferential direction of the flow of information. This fixed direction reflects the causality of the system modeled by the network. From the general perspective of the development of a computational tool, this unidirectional flow of information poses a significant constraint because it introduces an insensitivity with respect to future states of an independent variable. As an example, when the states are used to represent the time variable, the unidirectional flow does not pose an important limitation because causality must be respected. However, when the states are used to represent a spatial variable the unidirectionality can result in an insensitivity to boundary conditions.
This constraint on the flow of information was overcome by the more recent development of Bidirectional Recurrent Neural Networks (BRNN)~\cite{RNN-Review, BRNN-Schuster1997} that allow a two-way propagation of the information through the network (i.e. both forward and backward). We will show how this bidirectionality property is critical for the successful realization of a NN-based computational framework for physical systems.

In general, training RNNs is a challenging task due to the presence of internal state feedback~\cite{Goodfellow-et-al-2016} within the architecture. Successful training could become more challenging when a complex physical problem is modeled via these networks, particularly if accurate predictions are expected. However, the quality of the prediction can be greatly improved by using customized multi-objective loss functions during the network training phase. Although at the early stages of NN analysis custom loss functions were not commonly used, the idea of a multi-objective loss function has attracted considerable interest particularly in physics-based NN modeling~\cite{Physics-NN1, Physics-NN2, raissi2019physics}. This aspect will be discussed in the following sections. 

The remainder of this paper is organized as follows. First, we briefly introduce some fundamental concepts and properties of the BRNN architecture. Then, in \S~\ref{sec2} we present the general conceptual approach at the basis of the proposed computational framework. In order to facilitate the understanding of the methodology and to provide a quantitative example of the numerical performance, in \S~\ref{bar} and \ref{Results} we specialize the framework for application to a specific class of one-dimensional physical problems consisting in the static response of an elastic bar subject to a distributed axial load.

\section{Bidirectional recurrent neural network}\label{BRNN_sec1}

In this section, we briefly introduce the BRNN architecture \cite{Goodfellow-et-al-2016}, its basic components, and the role of key features within the context of the numerical solver.
The general structure of a BRNN is shown in Figure~\ref{BRNN1}. A typical BRNN is composed of two sets of RNN cells (labeled RNN$_{\text{F}}$ and RNN$_{\text{B}}$) each one processing the sequence of input data $\text{In}_i$ in opposite directions, that is forward (left to right) and backward (right to left). For each input sequence state $i$, the RNN cells use two input, that are $\text{In}_\text{i}$ and the internal hidden state (HS), in order to calculate the output from both the forward ($\text{F-Out}_\text{i}$) and backward ($\text{B-Out}_\text{i}$) RNNs path, and to update the internal HS for the next step $i+1$. Note that the internal HS provides information about the previous steps.
The output of the RNN cells in each direction is then combined to calculate the global output following each input state $1,2,...,n$. Among the several different properties of BRNNs, two stand out when analyzed in reference to the specific computational application proposed in this paper. These two properties are:
\begin{figure}[!ptb]
	\centering
	\includegraphics[width=.5\linewidth]{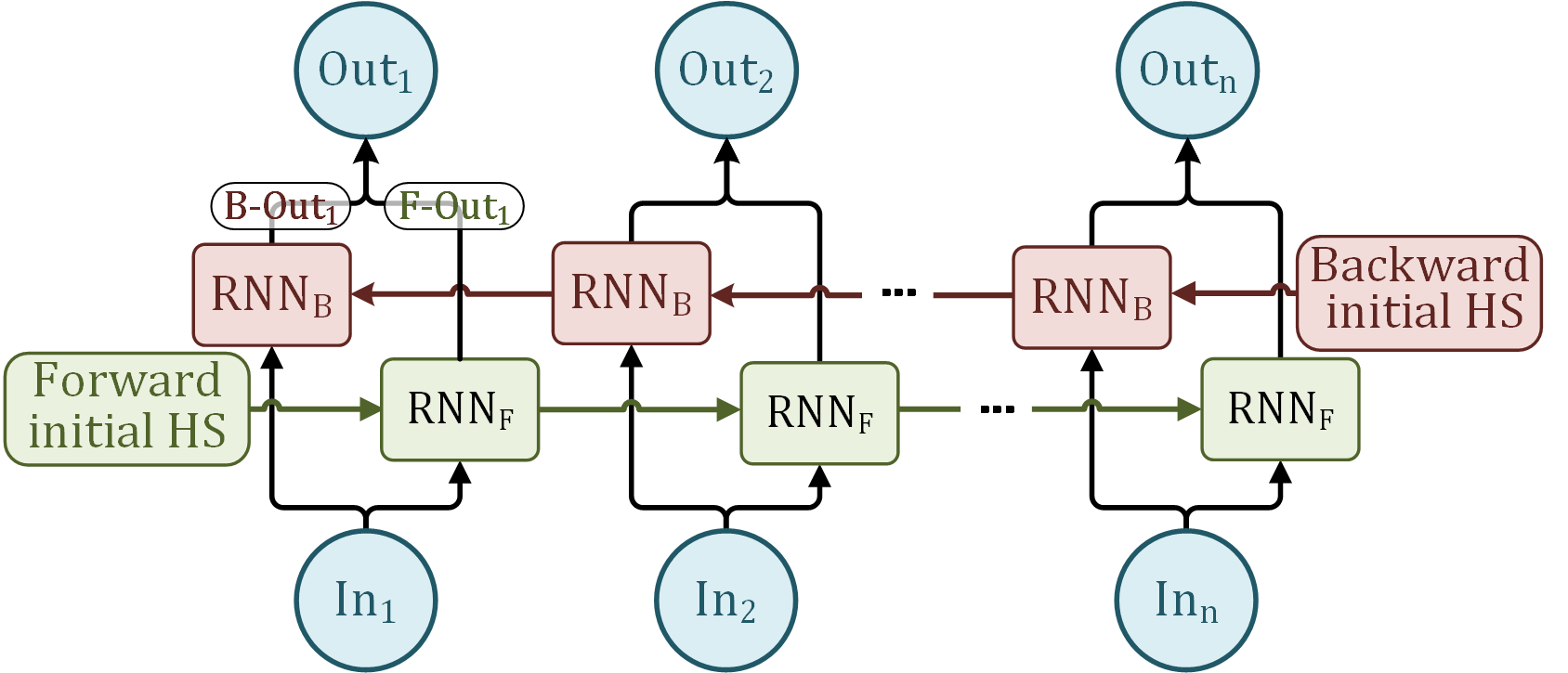}
	\caption{Expanded schematic of a bidirectional recurrent neural network (BRNN) showing its internal structure. Each element of the input sequence $\textit{In}_i,\ i =1,2, ... , n$ is processed in two opposite directions via the recurrent networks RNN\textsubscript{F} (forward) and RNN\textsubscript{B} (backward). At each state $i$, the output of these two recurrent networks is combined to calculate the output at the $i$-${th}$ state $Out_i$ as well as the hidden state. Therefore, both previous and future data in the input sequence are considered in output predictions. In each direction, the initial hidden state (HS) can also be defined as network input describing the initial or boundary conditions of the problem.} 
	\label{BRNN1}
\end{figure}
\begin{itemize}
	\item{ \underline{Variable input sequence size}: BRNN is able to learn the sequential logic behind an input-output sequence. Thus, for problem domains with different sizes (hence with variable input sequence size), BRNN is still capable of processing the input data and predict the output.}
	\item{ \underline{Bidirectionality}: the basic RNN determines the current state of the output only based on all the input prior to the current state (an implicit assumption of output sequence causality). 
		It follows that an output at a given 
		state cannot be affected by the future states of the input (or, equivalently, it does not depend on future events). While this observation is perfectly consistent with our understanding of causality in physical systems (that is when the states define a temporal sequence), this same condition turns into a very limiting constraint when the states represent an independent variable such as space. In this latter case, the unidirectional flow of information would result in an insensitivity with respect to certain input and physical boundaries. It follows that, in order to be amenable to the solution of physical problems defined in a spatio-temporal domain, the network must receive feedback from either previous or future states. 
		This feedback requirement is enabled by the concept of BRNN which allows two opposite paths of propagation of the information: forward and backward.
	}
\end{itemize}

\section{Finite Element Network Analysis (FENA): basic concepts and fundamental modeling strategy}\label{sec2}

\begin{figure*}[!htpb]
	\centering
	\includegraphics[width=.99\linewidth]{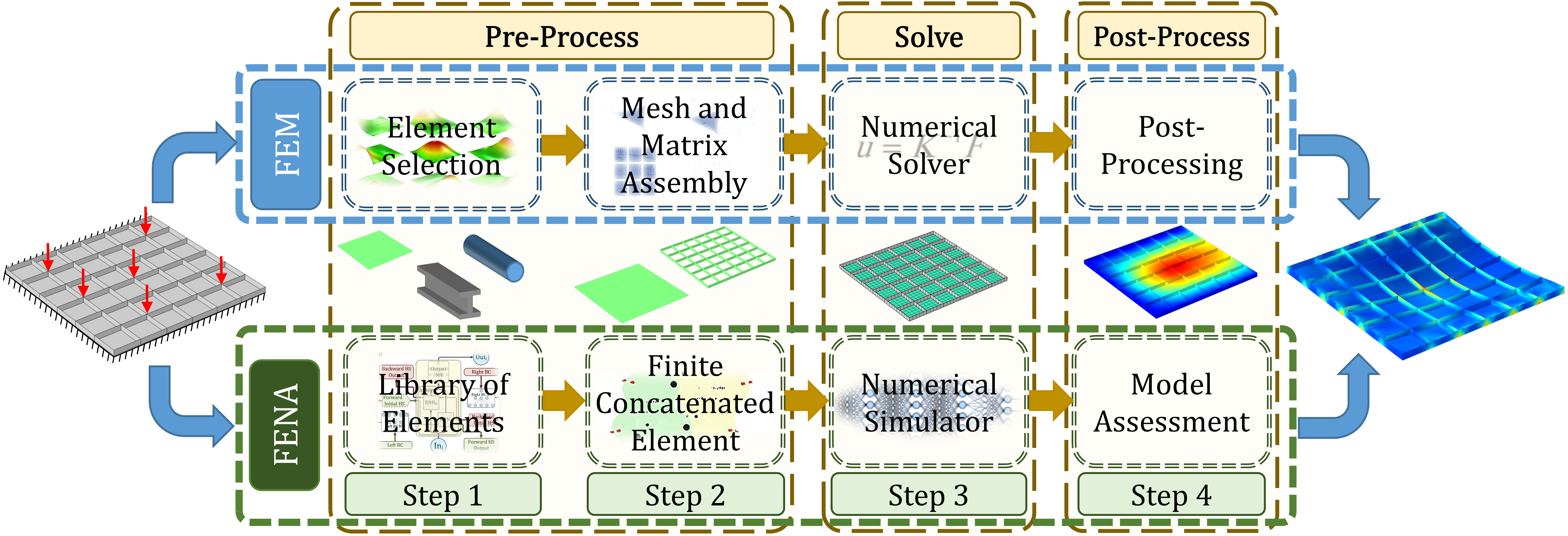}
	\caption{Conceptual view presenting the main modules of the finite element network analysis (FENA) framework and the logical sequence of operations to build a model and produce a numerical solution. The logical flow of operations is exemplified for the structural analysis of a stiffened panel under specified loads and boundary conditions. In order to facilitate the understanding of the FENA framework, the logical process is directly compared with the process required to assemble a similar computational model in a more traditional FE analysis (FEA) framework. Assuming a target problem consisting, for example, in the calculation of the static response of a stiffened panel under prescribed loads and boundary conditions, pre-trained network models for beam and plate elements are selected from the existing library LE (Step 1). If necessary, multiple network element can be concatenated to create complex assemblies of the fundamental elements (Step 2). Following Step 2, a network model of the complex structural assembly is available and ready for the calculation of the solution according to the selected physics (static response in this case). The network predicts the output in terms of the static displacement profile (Step 3), and any other output quantity the individual network elements in LE were designed to provide; for example stresses and strains. Finally, the MA module is used to assess the reliability and accuracy of the calculated solution (Step 4). 
	}
	\label{FENA-Schem} 
\end{figure*}

In this section, we introduce the general concept that enables building predictive computational models of physical systems based on interconnected pre-trained NN. In the following, this approach will be referred to as \textit{Finite Element Network Analysis} (FENA) in order to highlight one of the major characteristics of this method that is the ability to build complex models by combining existing (and pre-trained) neural networks mimicking specific physical behaviors. FENA builds upon four major components (or modules): a Library of Elements (LE) module, a Finite Concatenated Element (FCE) module, a Numerical Simulator (NS) module, and a Model Assessment (MA) module. A conceptual view of FENA's inner structure is provided in the schematic of Figure~\ref{FENA-Schem} while the role of the fundamental components is discussed hereafter. In an attempt to facilitate the understanding of this novel conceptual approach to physical simulations, Figure~\ref{FENA-Schem} compares the different steps in FENA to conceptually equivalent steps in a more classical finite element approach. From a high-level perspective, the four modules of FENA are identified as follows:

\begin{itemize}
	\item \underline{Library of Elements (LE):} LE is a database that includes an ensemble of pre-trained physics-informed BRNN, whose role is to simulate the response of selected physical components or systems. As previously mentioned, this paper will illustrate the computational methodology in the context of structural analysis, hence the library will focus on neural networks mimicking the behavior and response of structural elements. For the sake of this work, structural elements will be limited to 1D elastic bars. From a conceptual perspective, LE takes the role that element types and shape functions have in classical finite element analysis.
	\item \underline{Finite Concatenated Element (FCE) module:} the FCE module allows connecting multiple BRNN elements in order to assemble models of complex systems. This functionality serves two main roles: 1) it allows combining networks that simulate different parts of a complex physical system including different material and geometric properties, boundary conditions, even different physics, and 2) it provides a unique foundation to the efficient simulation of very large systems. This latter capability is somewhat reminiscent of dynamic condensation and sub-structuring approaches that are well-known in FEA\cite{craig2006fundamentals}.
	\item \underline{Numerical Simulator (NS) module:} this module executes the network in order to produce the actual numerical solution. All the problem-specific input (such as loads and boundary conditions) is applied at this stage. Then, the pre-trained network model is numerically executed to produce an estimate of the output variables. In terms of functionalities, NS is reminiscent of the numerical solution of the system of equations in the FE method. 
	\item \underline{Model Assessment (MA) module}: 
	once the response of the system is calculated by the NS module, the MA module provides a series of tools to assess the reliability and accuracy of the numerical results. 
	Recall that the network training is a probabilistic process due to both stochastic initialization and selection of the training batch. During training, weights are determined to minimize its loss function and maximize prediction accuracy for a given training data set. The main goal of this module is to assess the performance and the accuracy of the resulting network model and its ability to correctly simulate the response of a given system under different conditions (e.g. loads, materials, geometry, and constraints).
\end{itemize}

Given these four modules, the following four-step procedure (Figure~\ref{FENA-Schem}) should be implemented in order to build the model of a physical system and to perform numerical simulations in FENA:

\textbf{Step 1:} Select from the library LE the appropriate network elements to represent the various components of the physical system.  
Note that a BRNN network element does not map typically to a single finite element but instead it represents entire sub-components of the system.

\textbf{Step 2:} Depending on the properties and size of the physical domain to be simulated, the FCE module may be used in order to create complex assemblies (i.e. combinations of multiple sub-domains). FCE is used to combine sub-domains (i.e. different BRNN elements) when the computational domain is described by, for example, different physical behaviors, governing equations, geometric and material properties, and spatial scales. From a more practical perspective, the LE and FCE modules are applied concurrently to build the model of the physical system. This operation bears some similarity with the FE meshing process where the system is discretized and prepared for the numerical solution. 

\textbf{Step 3:} Once the equivalent network model is assembled, and provided the sequence of problem-specific input (e.g. input loads and boundary conditions), the network is executed in order to compute the output. This operation is conceptually equivalent to the numerical solution of the system of algebraic equations performed in FEA. However, it is this step that leads to the most significant differences in the required computational time and resources between the two methods. In fact, while the matrix inversion involved in the classical FE solution is one of the most expensive computational task, no matrix inversion is required in FENA. Besides, the evaluation of a trained network is an extremely fast and efficient process that requires only a fraction of the time needed by FEA. Quantitative details will be provided in the numerical example below.

\textbf{Step 4:} Once the output of the network (i.e. the response of the physical system) is calculated, the final step involves the use of the MA module in order to analyze the accuracy and reliability of the solution produced by the network. 

\section{FENA for structural analysis}\label{bar}

In this section, we discuss the technical details of the methodology by illustrating its application to the case of structural analysis. More specifically, we illustrate the general procedure to develop a deep learning based numerical analysis framework for structural analysis. In the following section, we will apply this framework to solve representative numerical problems and to discuss results and  performance.

We focus on implementing the methodology for the static analysis of a bar subject to distributed axial loads. The bar can have variable cross-section, different material properties and boundary conditions, as well as it can be subject to an arbitrary distribution of axial load. All these elements can be specified as input when building the model. A schematic view of the selected structure is presented in Figure~\ref{Probl-Schem}a and it is represented by an elastic bar with fixed boundary conditions (end sections highlighted in gray) and subject to a distributed axial load (black arrows). The bar is divided into segments, each with potentially different material and geometric characteristics.

Before proceeding with our discussion, it is worth highlighting a few aspects related to the selection of the benchmark model that was chosen as representative physical system for our analysis. While in the context of FEA the bar problem does not bear any significant complexity, in the context of neural networks the ability to
build a general NN-based computational methodology capable of addressing different geometric, material, loading and boundary conditions is a very challenging task. This process becomes even more demanding if multiple networks must be combined to form mechanical assemblies without any further need for training.  At the same time, many of these challenges are common to several classes of boundary value problems and physical systems and are not exclusively related to the problem dimensionality.
This study focuses on introducing a new concept of forward solver based on two features that, to the best of the author's knowledge, were never shown before in the literature: 1) the NN-based solution of boundary value problems without the need for re-training, and 2) the concatenation of NN elements (also without the need for re-training) to provide modularity in assembling systems from basic components. The selected 1D bar problem allows complete focus on the conceptual and numerical aspects of the methodology without clouding the description of the approach with considerations needed to support the analysis of higher dimensional elements.

In the following sections, we review individually the structure and the application of each FENA module as it pertains to the selected structural analysis problem.

\subsection{LE - network architecture design and training}\label{NetDesign}

The LE contains the fundamental set of elements that can be used to assemble the model of the physical system. In the present case, only elements mimicking the static response of bars are needed.

\begin{figure}[!ht]
	\centering
	\includegraphics[width=.45\linewidth]{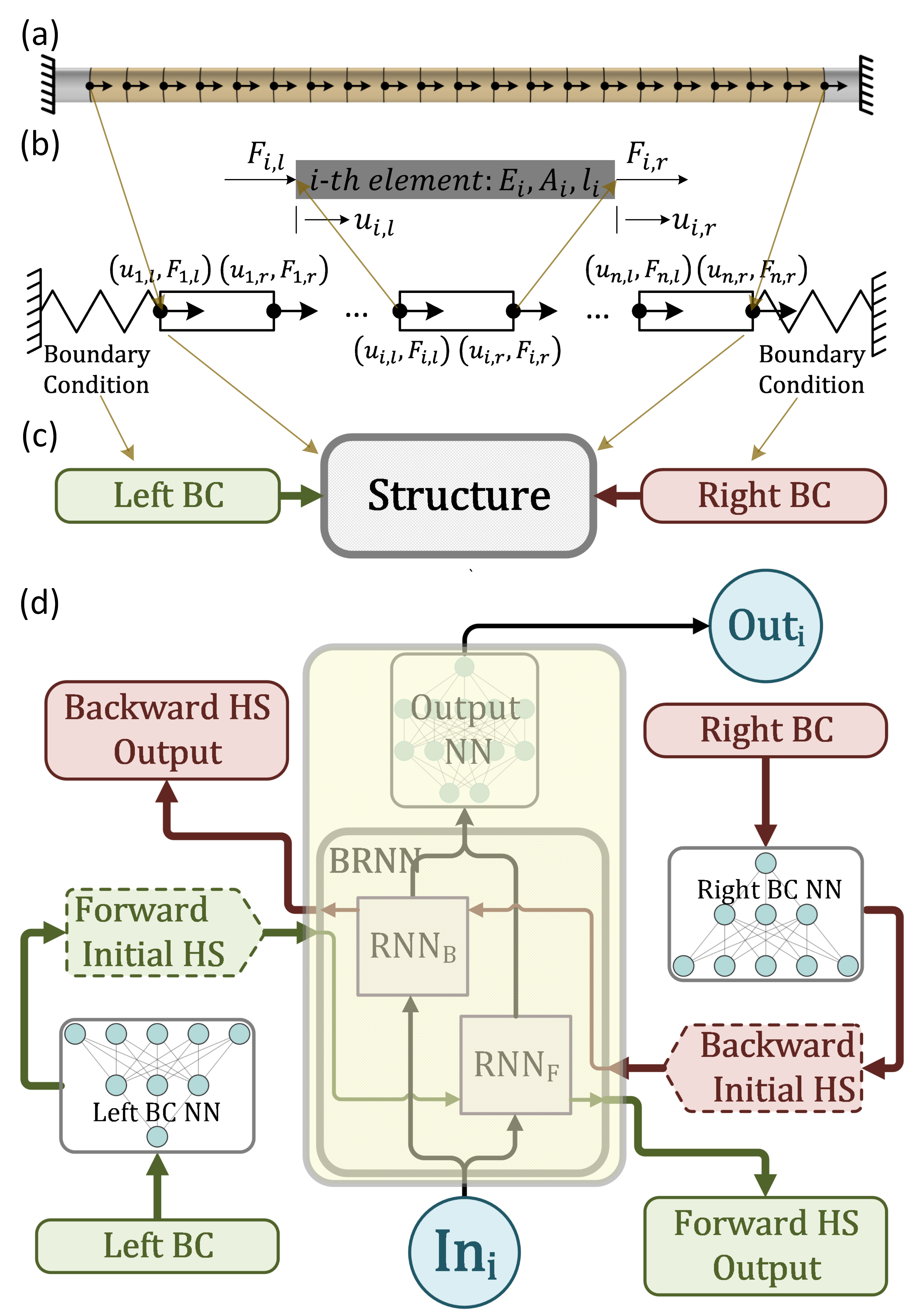}
	\caption{Schematic showing the implementation of a BRNN computational block exemplified for a structural elastic bar element. (a) Elastic bar showing the applied axial input loads (black arrows), the boundary conditions (gray elements), and a possible partitioning in elements. (b) A schematic view of the different elements indicating the displacement $u$ and force $F$ nodal variables. (c) High-level view of the network implementation corresponding to the bar. (d) the detailed view of the BRNN network structure used to simulate the elastic bar. The network receives a sequence of input necessary to define the different sections (i.e. cross-section area, Young's modulus, element length, and applied load) as well as boundary conditions (passed to the network as initial values of the HS) and predicts the nodal displacements $u_{i,l}$ and $u_{i,r}$ for each element. The network contains a core bidirectional layer to learn sequence dependencies as well as fully connected layers added to extend its prediction capacity.} 
	\label{Probl-Schem}
\end{figure}

To be able to simulate a general elastic bar via BRNN, we first divide the computational domain into $n$ elements (Figure~\ref{Probl-Schem}a). This operation is needed to prepare the input sequence for the BRNN, because the external input (in this case the axial load) can be applied only at the nodes of the BRNN (see \ref{netstructure}). Given that BRNN can learn the recursive logic behind input and output sequence regardless of the sequence size, this is a critical step to obtain a general solver which is not limited to a fixed size domain. Similar to meshing strategy in the FE method, the number of these elements are connected to the desired output prediction accuracy and to the location of (network) output nodes. In other terms, it controls the access to either input or output nodal quantities (e.g. external loads or displacements). Nevertheless, contrarily to the concept of FE mesh, the nodal response can be very accurate even when the discretization is very coarse (this is due to the fact that the calculation in NN is based on complex transfer functions between nodes). At the same time, as in FE analysis, a coarser discretization can still alias the response, therefore introducing significant error.

Considering these factors, the partitioning can be determined. The goal is to find the left and right endpoints (nodes) displacements $u_{i,{l}}$ and $u_{i,{r}}$, with $\ i = 1,2,..., n$ and $n$ is the number of elements of the problem. The response depends on applied nodal loads $F_{i,{l}}$ and $F_{i,{r}}$, the element material and geometric properties here considered isotropic and expressed only in terms of the Young's modulus $E_i$, the cross sectional area $A_i$, and the length $l_i$ (Figure~\ref{Probl-Schem}b). The BRNN representing the bar element receives as input sequence these known parameters $F_{i,{l}}, F_{i,{r}}, E_i, A_i, l_i$ and returns the nodal displacements $u_{i,{l}}, u_{i,{r}}$.

The BRNN input sequence is processed in a recursive manner in both directions and the corresponding data (analog to internal nodal loads in FEA) between adjacent elements are transferred via the internal hidden states of the forward and backward RNN. 

In addition to the applied external loads, the bar response depends also on the boundary conditions (Figure~\ref{Probl-Schem}c). Together with the external loads, this is one of the most challenging conditions to manage with network-based models. The main reason is that loads and boundary conditions are analysis-specific input that strongly affects the response of either the individual elements or the whole system. While passing different numerical values to a pre-trained network is a trivial operation, the application of a different number of loads and boundary conditions (or any change to their points of application) is a much more complicated task to manage (because it implies a change in the size and structure of the input sequence) and typically require a complete retraining of the network.

For a generic elastic bar BRNN, the elements cross-sectional area, Young's modulus, start and endpoint coordinates (which determines the element length $l_i$), boundary stiffness, and the applied loads form the network input. The input sequence is passed through three layers with [14,14,42] number of neurons before being processed by the forward and backward RNN cells contained in the BRNN layer. The left and right boundary conditions are first passed through two fully connected layers which, in turn, define the hidden states for both forward and backward directions (Figure~\ref{Probl-Schem}d) RNN.
The BRNN layer is composed of 30  long-short term memory (LSTM) cells \cite{Gers1999} in each direction (Detail of LSTM cells are presented in \ref{RNN_LSTM_dis}) and its output is connected to 5 fully connected layers, the first of which counts 50 neurons while the remaining four count 60 neurons. The specific size of the different parts of the network architecture was obtained via a trial and error procedure with particular attention to the numerical stability of the training process and to the overall accuracy of the prediction. Note that a similar network architecture could be used to model a bar with assigned displacement boundary conditions. More specifically, the displacement boundary conditions would be provided to the network in place of the boundary stiffness values.

In order to ensure that accurate predictions could be achieved for a broad range of output (i.e. ranges of amplitude of the nodal displacement), a tailored loss function was defined to include three equally weighted and normalized terms: 1) a mean square error of the predicted nodal displacement, 2) a nodal displacement continuity condition, and 3) a physical law. The first term is the commonly used cost function when solving regression problems via NN. The second term enforces the continuity of the physical displacement between adjacent nodes of two neighboring elements. The third term helps the network identifying solutions that satisfy the equilibrium equation for the specific physical system (in this case, and elastic bar). The networks are trained using ADAM algorithm \cite{adam} with a variable learning rate initialized at 0.0001, using a FE model generated training sets. \ref{netstructure} provides in depth details of the network architecture, the training data set parameters, and the training procedure.

\subsection{FCE - network assembly}\label{FCE}

The previous section described the procedure to build a library of predictive NN-based models capable of simulating the behavior of elastic bars. The networks are general enough to accept a variable-size sequence of input parameters defining both material and geometric properties, as well as external loads and boundary conditions. To some extent, these networks can be seen as surrogate models of the physical elements. The modeling of complex systems requires the ability to combine multiple dissimilar components (representing, for example, different physics or mechanical behavior) into a cohesive assembly. The ability to create these assemblies with NN-based models requires dedicated strategies allowing interfacing and integrating different networks. A successful network integration is also the key towards the analysis of large scale systems. In fact, the basic network discussed in \S~\ref{NetDesign} gradually loses accuracy as the domain total number of elements exceeds the range used in the training data set, hence it works well to simulate sections but it cannot be used by itself to simulate arbitrarily large systems. 

\begin{figure}[!hb]
	
	\centering
	\includegraphics[width=.7\linewidth]{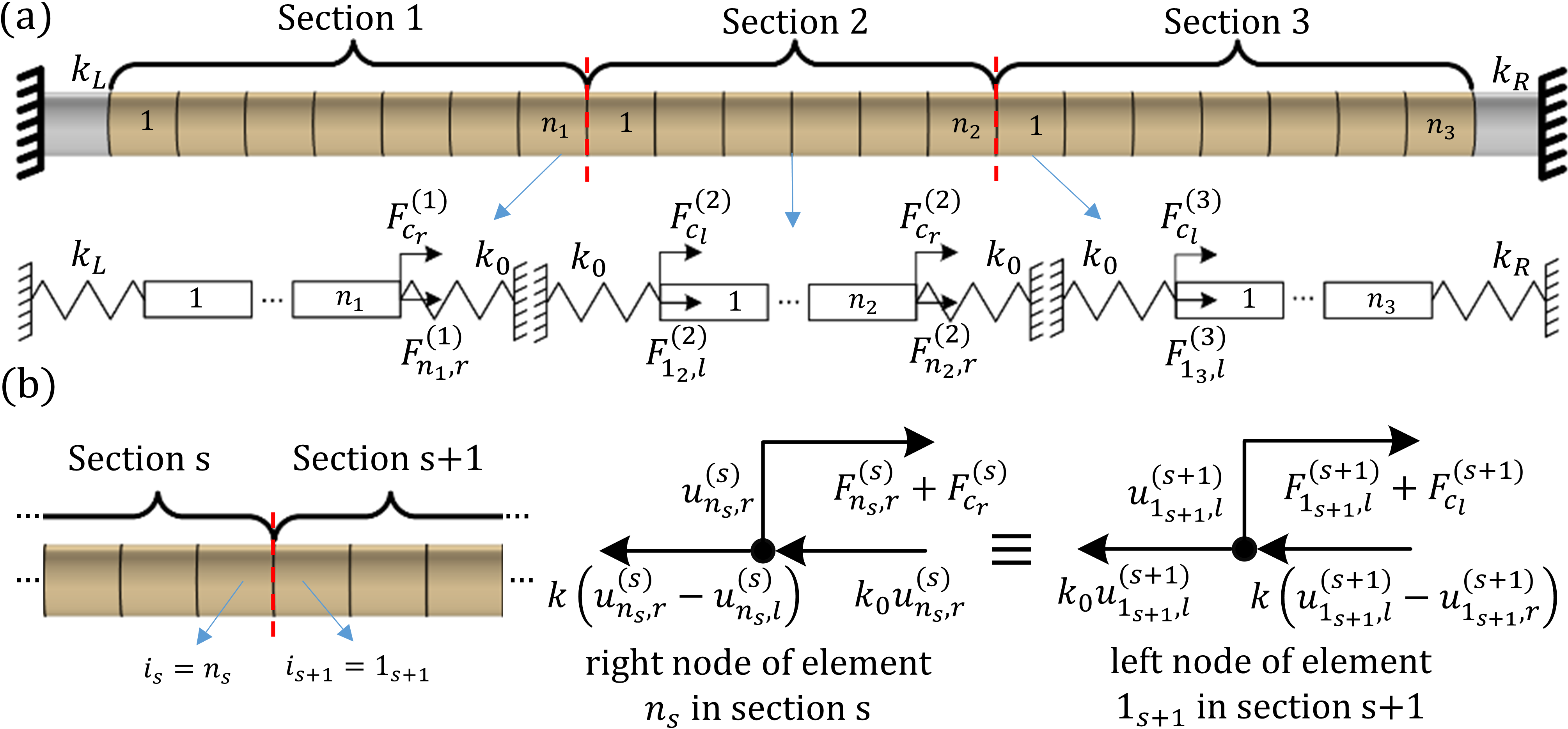}
	\caption{Schematic illustrating the concept of network concatenation (FCE) for the case of an elastic bar. (a) This example considers a bar of length $L$ obtained by combining three subsections of respective length $L_1,L_2,$ and $L_3$ each subdivided in $n_1,n_2,$ and $n_3$ total number of elements. The continuity conditions between adjacent network elements is obtained via a corrective load strategy $F_c$. (b) Schematic view indicating the main displacement and load variables for the interface between subsections $s$ and $s+1$. $k$ and $k_0$ are the element and the virtual boundary stiffness. The corrective forces $F_{c_r}^{(s)}$ and $F_{c_l}^{(s+1)}$ allows implementing the continuity conditions between adjacent network elements when forming the assembly.
	}\label{FCE_single_pic}
	
\end{figure}

To overcome this important issue, we propose the concept of network concatenation, which is schematically illustrated in Figure~\ref{FCE_single_pic}a for the case of a three-section bar structure. Initially, each section is considered as a standalone element with its own boundary conditions; essentially each element corresponds to the fundamental network available in LE, as described in \S~\ref{NetDesign}. The element boundaries are indicated by a finite stiffness of generic value $k_0$ and can be thought as virtual boundaries because they will be used to implement either boundary or continuity conditions depending on the position of each element within the concatenated structure. To clarify this point, in the example of Figure~\ref{FCE_single_pic}a the left virtual boundary of section 1 will be used to set a true boundary condition, while the right virtual boundary of section 1 will be used to set a continuity condition with the adjacent section 2. 

In order to achieve efficient concatenation of the networks selected from LE, we adopt a corrective load strategy. The corrective load $F_c$ is a virtual load that is added to the boundary node of each section to enforce continuity. In practice, the corrective load mimics the internal forces that arise from the continuity of adjacent sections. However, $F_c$ is not exactly coincident with the internal load because it also accounts for the effect of the virtual boundary stiffness $k_0$. It follows that boundary nodes on each subsection are under the combined effect of both externally applied and corrective loads.

Figure~\ref{FCE_single_pic}b shows a conceptual schematic of the different sections being connected at boundary nodes and marks the corresponding corrective loads. The corrective load at a generic interface (Figure~\ref{FCE_single_pic}) connecting sections $s$ and $s+1$ can be estimated imposing a classical Newtonian equilibrium. The equilibrium is given by:
\begin{equation}\label{FCE-1}
	\begin{aligned}
		& F_{c_{r}}^{(s)} = k(u_{n_s,r}^{(s)}-u_{n_s,l}^{(s)})+ k_0 u_{n_s,r}^{(s)} - F^{(s)}_{n_s,r}\\ 
		& F_{c_{l}}^{(s+1)}= k(u_{1_{s+1},l}^{(s+1)}-u_{1_{s+1},r}^{(s+1)})+ k_0 u_{1_{s+1},l}^{(s+1)}- F^{(s+1)}_{1_{s+1},l}
	\end{aligned}
\end{equation}     

\noindent where the superscripts $(s)$ and $(s+1)$ indicate the section number, while the two subscripts indicate the element number within each section (e.g. $n_s$ indicate the $n$-th element in section s, and $1_{s+1}$ indicate the first element in section $s+1$) and the node of application within that element (either $l=left$ or $r=right$).
Also, $k$ is the element stiffness that while formally should have the same indicial format $k^\Box_{\Box,\Box}$ we assumed as a constant for all sections in this example. Further, $k_0$ is the virtual arbitrary boundary stiffness, $F_{n_s,r}^{(s)} = F^{(s+1)}_{1_{s+1},l}$ is the external load at the interface between section $s$ and $s+1$, $F_{c_{r}}^{(s)}$ and $F_{c_{l}}^{(s+1)}$ are the corrective loads at the same interface, and $u^\Box_{\Box,\Box}$ is the nodal displacement. When network elements are concatenated, the corresponding corrective loads at the interfaces are estimated via an optimization process. For a domain divided into a total of $S$ sections (with $S-1$ interfaces) the optimization cost function $\mathbf{J}$ is defined as:
\begin{equation}\label{FCE-2}
	\begin{aligned}
		&\mathbf{J} = a\times\mathbf{J}_1 +b \times \mathbf{J}_2 + c\times\mathbf{J}_3 \\ 
		&\mathbf{J_1} = \bar{k}^2\sum_{s=1}^{S-1}{\left(  u^{(s)}_{n_s,r} - u^{(s+1)}_{1_{s+1},l}\right) ^2}, \ 
		\mathbf{J_2} = \sum_{s=1}^{S-1}{\left(EA \frac{\partial u^{(s)}_{n_{s,r}}}{\partial x} -  EA \frac{\partial u^{(s+1)}_{1_{s+1,l}}}{\partial x} -F^{(s)}_{n_{s},r}  \right)}^2\\   
		&\mathbf{J_3} = \sum_{s=1}^{S}{\left( u^{(s)}_{1_{s},l} k + u^{(s)}_{n_s,r} k - \sum_{\substack{j\in\text{ nodes} \\ \text{of section s}}}^{} {F_j}  \right)}^2  
	\end{aligned}
\end{equation}
where $a,b$, and $c$ are the weighting factors, $\mathbf{J}_1$ and $\mathbf{J}_2$ enforce continuity of the displacement and forces at the section interfaces, while $\mathbf{J}_3$ imposes the equilibrium of the entire system as a whole. In $\mathbf{J}_1$, the coefficient $\bar{k}$ is the average element stiffness that is added for dimensional consistency. The corrective loads do not appear explicitly in the cost function $\mathbf{J}$, however they enter the problem via their effect on the displacement field at each section (Figure \ref{FCE_single_pic}). To solve for the value of the corrective load $F_c$, we apply the Newton Conjugate Gradient (Newton-CG) method, which was selected for its fast convergence rate. It is worth mentioning that this part of the process could be replaced by other possibly more efficient approaches and it is not limited to the Newton-CG method.    

\subsection{Accuracy assessment}\label{MA-description-BAR}

The present method uses neural networks to predict the response of physical systems. As discussed in \S~\ref{sec2},
dedicated metrics need to be developed to assess both accuracy and reliability of the predictions. The availability of these metrics is particularly important when the networks are used for predictive purposes, such as in the case of FENA. These metrics must indicate the level of confidence in a given prediction. In the following, we introduce four evaluation metrics and discuss their relevance and use in the context of FENA. The four metrics are: 1) MC index, 2) equilibrium satisfaction index (ESI), 3) linearity index, and 4) similarity index (model ensemble). The first metric is supposed to be used when building the surrogate models to populate the LE database. This metric provides a preliminary estimate of the expected accuracy for each pre-trained network in LE. The remaining metrics are applied to estimate the accuracy of solutions calculated by FENA.

The first metric is based on the Monte Carlo (MC) sampling method \cite{MC}. This metric takes sample simulations from the trained network (by solving multiple sample problems subject to randomly defined loading and boundary conditions) and calculates the expected mean relative error and its standard deviation. The relative percentage error at the $i$-${th}$ element is defined as $ \left( u^{pred}_{i,j} - u^{true}_{i,j} \right)/ max(u^{true}_{\Box}) \times 100, \ j= l,r$, where $u^{pred}_i$ is the network prediction, $u^{true}_i$ is the exact solution, and   $max(u^{true}_{\Box})$ is the maximum nodal displacement encountered in the simulation. The expected accuracy can be classified based on the order of magnitude of the output. The lower the expected relative percentage error, the higher the accuracy of network predictions. Note that this metric requires either a fast algorithm for estimating the ground truth (i.e. $u^{true}$) given the sampled input parameters or a very large dataset (which may not be available in complex practical applications).

The second index builds upon the existence of governing equations describing the specific physical problem. 
An Equilibrium Satisfaction Index (ESI) can be defined as:
\begin{equation}\label{equilib-eq}
	ESI = \frac{\frac{1}{n} \sum_{i=1}^{n} { \left( \begin{aligned} 
				F_{i,r} - &\underbrace{ (u_{i,r} - u_{i,l}) \times k_i}_{\text{Internal load at element i}}\\ +&\underbrace{(u_{{i+1},r} - u_{{i+1},l})\times k_{i+1}}_{\text{Internal load at element i+1}} 
			\end{aligned} \right) ^2}}{  \frac{1}{n}\sum_{i=1}^{n}{F_{i,r}^2}} 
\end{equation}

Equation~\ref{equilib-eq} is built around the concept of Newtonian equilibrium (enforced at all nodes) and it estimates the error in satisfying the governing equations. 
A zero value of the ESI would indicate the predicted and exact solution coincide. 
Although this metric provide an estimate of the goodness of the solution, there is no unique approach to define threshold values to determine how good a solution is, hence \textit{ad hoc} threshold should be set based on the physical understanding of the problem. Also, additional metrics should be used in conjunction with ESI to estimate the accuracy of the solution. It follows that this metric is recommended to be used to discard network predictions but not to assess the overall accuracy.

The third metric is the linearity index, which utilizes expert knowledge on certain characteristics of the system (in this case, the linear behavior) to evaluate the overall network performance to a given input sequence. In linear systems, correlated output is obtained following scaled input sequences. Using the Pearson product-moment correlation coefficient (PPMCC), which is a scale-free measure of the linear correlation (i.e. of the similarity), this index verifies if predicted network output sequences at multiple scaled input are correlated. The closer this index is to unity (for multiple solutions), the more reliable the output predictions. This index can also be used to detect a threshold on values of the input beyond which the network output starts deviating and accumulating error. It is worth mentioning that the basis of the linearity index is the concept that only a properly trained and accurate network returns correlated output for scaled input. If, for a given input sequence, the network accuracy is poor then it is highly probable that the predictions are uncorrelated due to the non-linearity of the network. In the case of the static linear elastic response of a bar, the transfer function between the output nodal displacement and the input force is linear. In order to use the linearity index, we calculate the network output for input loads scaled by different factors. These factors should be chosen in a way that the scaled nodal loads remain in the same range as those defined for the network training. The PPMCC matrix is then calculated for all the possible output solutions obtained for different scaling factors. If the PPMCC is close to unity, it is expected that the same displacement profile is obtained for all the scaled input.

The last metric employs an ensemble of neural networks to analyze the prediction accuracy by assessing similarity in their responses. It is only applicable when multiple trained network models are available for a problem. It is known that an ensemble of surrogate models can be used to improve the overall quality of the prediction~\cite{ensembleofnets}. As an example, in the case of neural networks applied to regression problems, different networks can be simultaneously trained and the average of their respective predictions is used to predict the overall output (this process is similar to voting in classification applications~\cite{battiti1994democracy}). Special care must be taken when using network ensemble averaging because a few highly erroneous predictions can severely compromise the whole average. In general, it is more appropriate to inspect first the predicted output from individual networks (for instance comparing them using the PPMCC) and then to selectively decide the networks to be retained for the averaging process. 

Note that in a linear problem the linearity index and the model ensemble method can be combined. To evaluate the quality of the predicted response with a combined index, first a problem is simulated via multiple networks and the PPMCC matrix of the responses is formed. If all the results are highly correlated the average prediction is accepted as the problem solution. If the solutions are not correlated, we keep scaling the input load and recalculate the output to obtain a scaled solution at which all the predictions by the network models in the ensemble are highly correlated. We will apply this combined method on the sample problems of \S~\ref{Results}.

\section{Static analysis of an elastic bar}\label{Results}

In this section, we present and discuss the application of FENA to a practical problem consisting in the solution of the static response of an elastic bar under a distributed axial load. The intent of this section is to demonstrate the potential of FENA for structural simulations and to provide an assessment of its performance. 

We consider the general problem of the static response of a homogeneous uniform bar with finite stiffness boundaries at both ends, under an external distributed axial load. Four sample cases will be considered and their properties are presented in Table~\ref{table:sample-case-prop}. The area, Young's modulus, and total number of elements are arbitrarily chosen from a uniform distribution within the ranges used to generate the training data sets (see \ref{Training}).

\begin{table}[!ht]
	\centering
	\resizebox{.7\columnwidth}{!}{
		\begin{tabular}{lrrrr} 
			&Case 1& Case 2&Case 3& Case 4\\ 
			~ & (Fig.~\ref{R1}a) & (Fig.~\ref{R1}b) & (Fig.~\ref{FCE_results}a) &
			(Fig.~\ref{Poor_pred}) \\
			\hline
			E [kPa]& $176.0$ & $97.7$ & $100.0$ &
			$97.4$ \\ 
			A [m\textsuperscript{2}]& $\medmuskip=0mu 1.6 \times 10^{-4}$ & $\medmuskip=0mu 8.6\times 10^{-5}$ & $\medmuskip=0mu 10^{-4}$ & $\medmuskip=0mu 5.1\times 10^{-5}$ \\ 
			left boundary stiffness [N/m] & $104$ & $961$ & $697$ & $698$ \\ 
			right boundary stiffness [N/m]& $1120$ & $1400$ & $1165$ & $1170$ \\ 
			L [m] & $3.60$ & $5.70$ & $24$ & $6.43$   \\
			n (number of elements) & $27$ & $45$ & $240$ & $45$   \\ 
			\hline
		\end{tabular}
	}
	\caption{Numerical values of material and geometric parameters selected for the four sample cases. The most significant differences between these four cases consist in the different total length of the bar and in the number of total network elements used.}
	\label{table:sample-case-prop}
	
\end{table}

The four cases were selected in the following way. Case 1 and 2 are meant to illustrate in detail the implementation of the solution procedure and that accurate structural mechanics solutions can be achieved via FENA. The main difference between these two cases is their spatial distribution of the applied load, that is arbitrary in case 1 and sinusoidal in case 2. Also, the two cases employ different domain parameters in order to show the accuracy of the pre-trained models under different problem conditions. Both cases can be simulated using a single BRNN network element, hence not requiring the application of the FCE module. Case 3 is conceived to illustrate the application of the FCE method on large scale structures. Note that the term large scale is used here to indicate any system whose number of elements exceeds the maximum number of elements set for the BRNN network elements during the training phase. Finally, case 4 focuses on the implementation of the MA module in a case where unreliable results are predicted by the network. This latter problem was specifically selected to show how model assessment strategies should be used to assess both reliability and accuracy of the solution and, in case of unreliable results, to adjust the problem parameters for better predictions. The remainder of this section will describe the specific approach to structural analysis. 

\subsection{Building LE for elastic bar elements}
In order to simulate the response of a bar under a static axial load, we built and trained four network models (indicated below by the label M\#). All the models used the same network architecture (see \S~\ref{bar}) but were trained under different conditions as described in Table~\ref{table:model_description}.
\begin{table}[!ht]
	\centering
	\resizebox{.6\columnwidth}{!}{
		\begin{tabular}{lcr} 
			Model name &Training data set No.& Loss function\\ 
			\hline
			M1& \#1&  No physical law\\ 
			M2& \#1&  physical-law informed\\ 
			M3& \#2&  No physical law\\ 
			M4& \#2&  physical-law informed\\ 
			\hline
		\end{tabular}
	}
	\caption{Four different network element models were developed to simulate the response of an elastic bar. These models formed the available LE for the numerical example. The four models differ based on the data set used for training and on the specific form of the loss functions, as reported in the table.}
	\label{table:model_description}
\end{table}

All of the four models were trained using the same training hyper-parameters (complete details on networks architecture, loss function formulation, training and validation data set generation, and the models training results are provided in \ref{netstructure} and \ref{sec:supplementary-results}).

The quality of the four different models can be assessed in terms of their ability to capture the structural response of the bar elements. As discussed in \S~\ref{MA-description-BAR}, this assessment is performed via the MA module. Once the basic element networks are trained, the MC index can be applied to estimate both the performance and expected error for each network. Note that this index is applicable since, at this stage, we assume the exact solution to be available via another methodology (i.e. a commercial FE solver). This assumption was made only to demonstrate, in this initial step, the ability of the network models selected for LE to simulate the physical response of the structural systems. Once the LE is available and validated, no reference solution is needed.

\begin{figure}[!th]
	\centering
	\includegraphics[width=.8\linewidth]{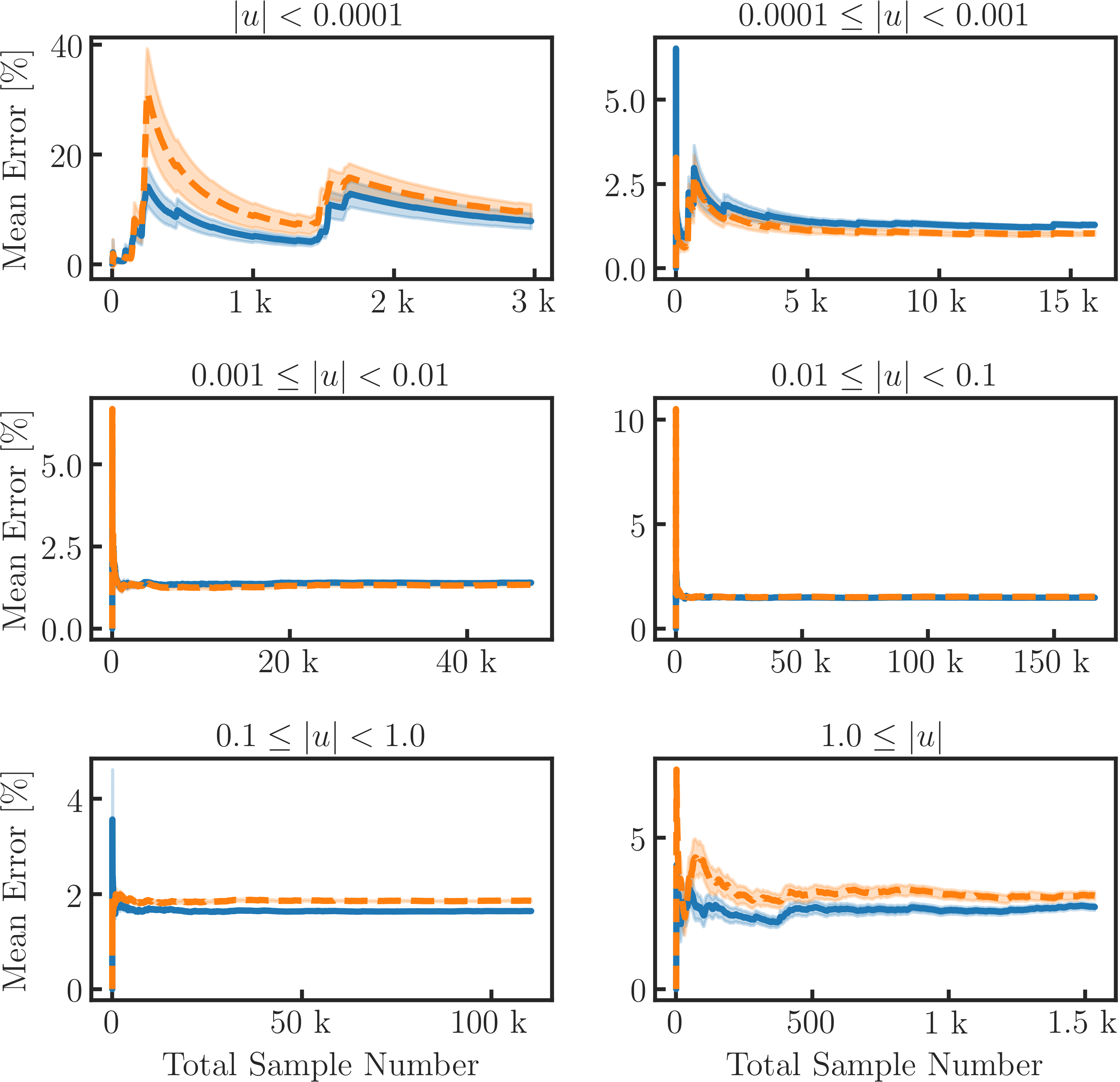}
	\caption{Sampling error obtained via the MC index for Model M1 (solid line) and model M2 (dashed line). The total sample number is the total number of nodes, across all the sampled problems, whose displacement is within the specified category. The shaded areas around each curve represent the 95\% confidence bands. The error plots are categorized by the range of displacement amplitude as reported in the label above each figure.} \label{MCfig}
\end{figure}

To demonstrate the procedure, we applied the MA module to models M1 and M2. Ten thousand sample problems were defined by assuming uniform distribution for each input parameter in the selected ranges. The percentage error on the predicted nodal displacement field was calculated and divided into six different categories (ranging from $10^{-4}$ to $1$) based one the order of magnitude of $|u|$. For each category, the expected value of the percentage error and its variance (95\% confidence bands, see shaded area) were calculated using the MC method (Figure~\ref{MCfig}). It is seen that any predicted nodal displacement with an absolute value greater than $10^{-4}$ is expected to have less than 2.5\% error. We note that the expected error value did not converge for displacement ranges below $10^{-4}$, which implicates unreliability of results within that range. We highlight that numerical cases involving displacement ranges smaller than $10^{-4}$ can be solved by properly scaling the input so that the resulting displacement values fall within the reliable prediction regions. 
The procedure discussed above leads to the formation of the LE with an established record of performance and reliability for the available elements. At this stage, the elements are available and ready to be used to build actual structural models. Note that, for future practical application and deployment of the FENA platform, the LE for a selected class of physical problems is expected to be previously built and available for use (hence taking an equivalent role to the library of elements and shape functions in the FE method).

\subsection{Case 1 and 2: static response under distributed loads}
In this section, we discuss the procedure to build the model and simulate the static response of the elastic bar under the conditions of cases 1 and 2. \\
\textbf{Pre-processing}: the first step consists in assembling the network model representative of the physical system. Recall that case 1 and 2 consider a bar whose length can be represented by a single network, hence not requiring any concatenation via the FCE module. The numerical values of the parameters for case 1 and 2 were provided in Table~ \ref{table:sample-case-prop}. To this end, BRNN networks are selected from LE based on the necessary range of parameters that are deemed appropriate for the simulation. 

\begin{figure}[!ht]
	\centering
	\includegraphics[width=.7\columnwidth]{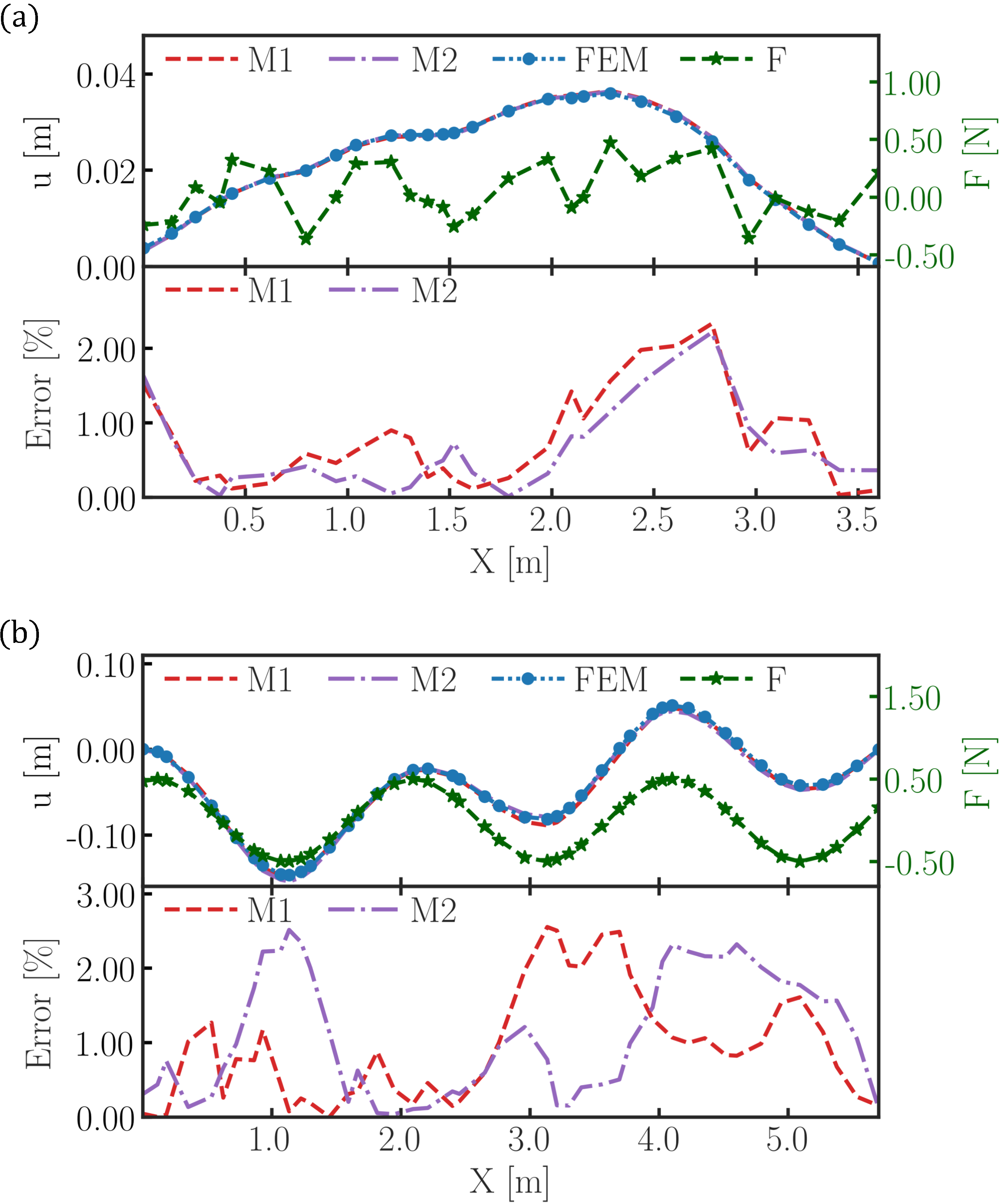}
	\caption{Numerical results representing the static response of the elastic bar under the conditions of cases 1 and 2. Nodal locations are identified by markers. The applied forced profile is indicated by the curve $F$. (a) Results for case 1 obtained by using both models M1 and M2 and a converged FE model (FEM). (top) displacement field profile, and (bottom) percentage relative nodal error calculated with respect to the FE solution. (b) Results for case 2 obtained by using both models M1 and M2 and a converged FE model (FEM). (top) displacement field profile, and (bottom) percentage relative nodal error calculated with respect to the FE solution.} \label{R1}
\end{figure}

In both configurations, the bar is subject to an axial distributed load as shown by the green curve (star markers with dashed line) in Figure~\ref{R1}. In case 1, the axial force is applied at 28 randomly distributed points within the bar length (Figure~\ref{R1}) and in case 2 it is applied at 46 points. The amplitude of the load is also generated randomly at these locations. Similar to the FE method, the external load is always applied in correspondence to a node (in this case a node within the BRNN unwrapped structure). This spacing between the loads affects the meshing (in FENA the sectioning) size. For case 1 and 2, the distance between each point of application of the nodal loads is within the range of element length used for training in all the four models types (M1-M4). Therefore, only one network element can be used to represent the whole system. In other terms, the FCE module is not needed for case 1 and 2. Note that, if either the total bar length or the total number of elements needed to solve a problem exceeds the ranges of the elements in LE, then more network elements must be combined together, hence requiring the FCE module. \\
\textbf{Numerical solution}: the previous step provides a network model that represents the physical system at hand.
The model is now ready to be passed to the NS module which prepares the input sequence (including cross-section areas, Young's moduli, length of each section, and nodal load), apply them to the network, and evaluate the network in order to simulate the response. Case 1 and 2 were solved using all four models. The distribution of input load as well as the numerical predictions from M1 and M2 are shown in Figure~\ref{R1}a-b (See \S~\ref{sec: M3-4Suppresults} for the results of M3 and M4). Both models provide a very accurate prediction of the static response of the elastic bar with a maximum error less than 2.5\% at each node (consistent with the assessment of the general performance of the elements in LE). The error is calculated with respect to a reference solution obtained via finite element analysis (performed via an in-house FE model code).\\
\textbf{Post-processing:} The final step of the simulation requires an assessment of the models' predictions. Clearly, in these sample problems this step appears to be redundant because we have access to a reference solution that serves as ground truth. However, a reference solution is typically not available hence the reliability and accuracy of the results must be assessed independently. Note that this assessment procedure can be thought as an operation equivalent to convergence analysis in the FE method, wherein different mesh sizes are tested to ensure the FE model results are accurate. According to the assessment strategy outlined in \S~\ref{MA-description-BAR}, we can calculate the ESI index. For case 1, it is found that $ESI_{M1}=0.098$ and $ESI_{M2}=0.023$ for model M1 and M2, respectively. In case 2, this index is equal to $ESI_{M1}=0.096$ and $ESI_{M2}=0.060$. Low values of the ESI index highlight limited reliability of the results. However, as previously mentioned in \S~\ref{MA-description-BAR}, the ESI index cannot be used independently in order to accept or discard solutions. Therefore, we solve the problem with all the available four models and calculate the similarity index. The PPMCC matrices for these two problems are equal to: 
\begin{equation}\label{PPMCC-P12}
	\begin{aligned}
		& \text{Case 1:}
		\begin{blockarray}{ccccc}
			\text{M1} &\text{M2} &\text{M3} &\text{M4} & \\
			\begin{block}{[cccc]c}	\bigstrut[t]	
				1&0.999&0.999&0.999&\text{M1}\\
				&1   &0.999&0.999 &\text{M2} \\
				&	  &1    &0.998 &\text{M3} \\
				\multicolumn{2}{c}{\text{\smash{\raisebox{1.5ex}{sym.}}}}&
				&1   & \text{M4} \bigstrut[b]\\
			\end{block}
		\end{blockarray}	
		& \text{Case 2:}
		\begin{blockarray}{ccccc}
			\text{M1}&\text{M2}&\text{M3}&\text{M4} &\\
			\begin{block}{[cccc]c}\bigstrut[t]
				1&0.997&0.993&0.998&\text{M1}\\
				&1   &0.997&0.998&\text{M2}\\
				&    &1    &0.994&\text{M3}\\
				\multicolumn{2}{c}{\text{\smash{\raisebox{1.5ex}{sym.}}}}&
				& 1   & \text{M4} \bigstrut[b] \\
			\end{block}
		\end{blockarray}
	\end{aligned}
\end{equation}
Different rows and columns are labeled with the corresponding model names and each element in the matrices represents the correlation coefficient between the selected model pairs. Results indicate that the output from different models correlate with each other within a $0.7\%$ error. This high level of correlation of the results indicates good reliability and accuracy of the predictions. 

\subsection{Case 3: element network assembly via FCE methodology} 
\begin{figure}[!ht]
	\centering
	\includegraphics[width=.7\linewidth]{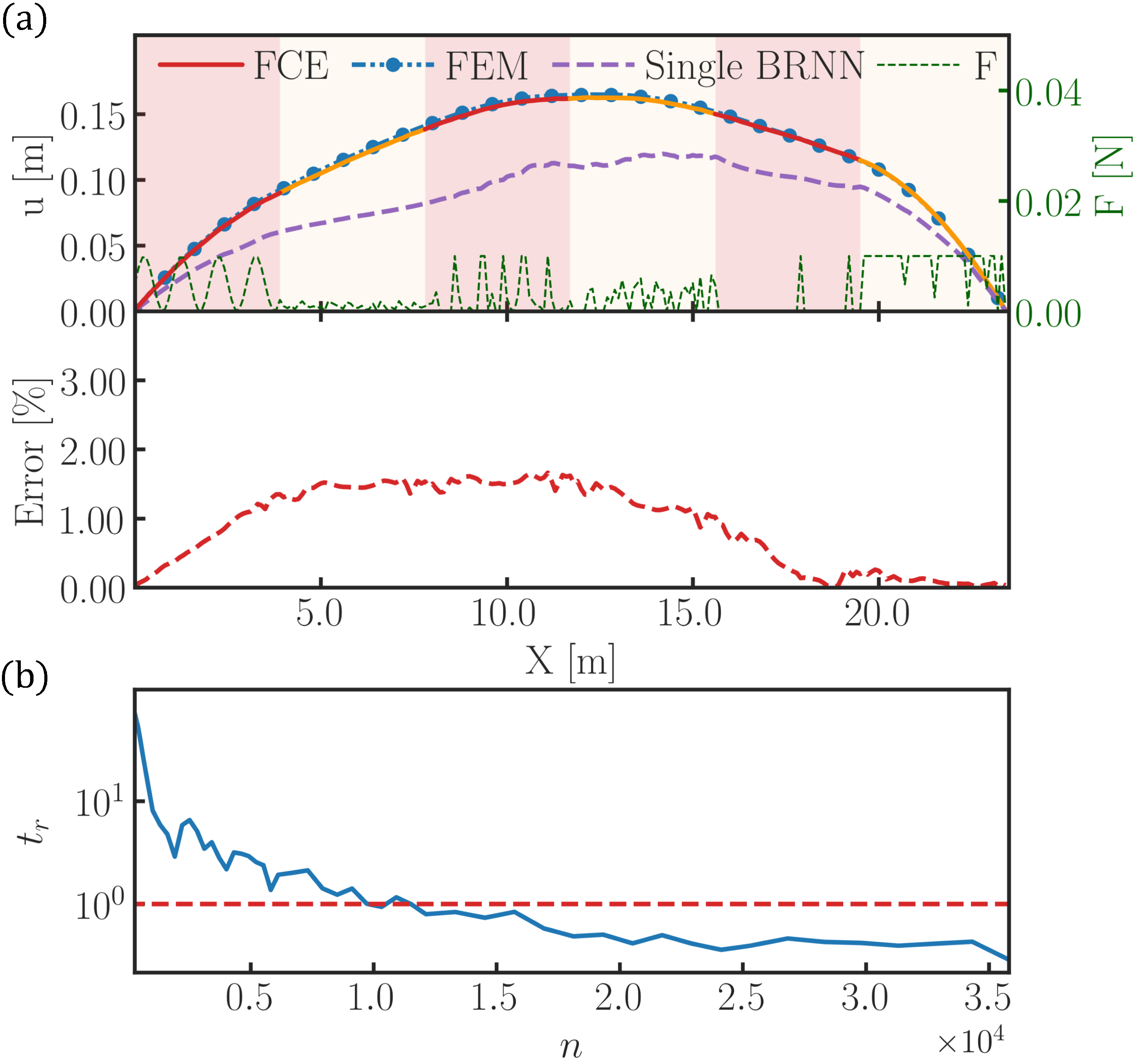}
	\caption{(a) (top) Numerical results in terms of the displacement profile obtained via FENA-FCE (solid line) using six sections and a total of 240 elements. Different sections in the FCE solutions are represented with different color shaded areas. The results from a converged FE model (FEM) was also reported as reference. (bottom) Percentage relative error calculated with respected to the FE solution. (b) Ratio of the computational time of FENA-FCE and of the FE method as a function of increasing number of elements $n$.}
	\label{FCE_results}
\end{figure}

In this section, we apply FENA to the solution of case 3 which requires the application of the FCE module. As summarized in Table~\ref{table:sample-case-prop}, case 3 considers a bar having a length $L=24\ [m]$ and subject to a distributed axial load applied every $0.1 \ [m]$. The specific external load being applied at each section is arbitrarily chosen and it is shown in Figure~\ref{FCE_results}a. The discretization of the load requires 240 elements which exceeds by a factor of three the maximum number of elements in the training samples. Thus, in order to build the model of this specific structure the FCE algorithm must be employed. 

To assemble the model, the domain was divided into six sections containing $[n_1,n_2 ...,n_6]=[40,40,40,$ $40,40,40]$ elements, respectively. The number of elements in each section was chosen to be $n_s \leq 60$, that is less than the maximum element size utilized during the training phase. Note that, at this stage, the choice of sectioning was arbitrary and several alternative options could be envisioned without exceeding the maximum element number $n_s=60$. 

Each of the six sections was independently represented by a dedicated BRNN receiving the input load corresponding to that specific section of the full-length structure (Figure~\ref{FCE_single_pic}). These individual sections were linked together via the FCE algorithm. 

In order to connect the six sections, ten (virtual) corrective loads were needed (two loads per interface). Based on the formulations presented in \S~\ref{FCE}, a cost function consisting of a weighted summation of the three terms given by Eq.~\ref{FCE-2} was assembled and a minimization procedure was used to determine the corrective loads to guarantee continuity of the sections. The weight factors of the cost function terms were selected as $a=.25$, $b=1$, and $c=.125$. The specific values of the weights were obtained by a combination of physical considerations and a trial and error process that showed how accurate results could be obtained over a wide range of problem sizes. The weights can be set when building LE and do not need to be determined again for a specific analysis. The minimization procedure was performed via the Newton-CG method (using python Scipy package) and converged in eight steps. 
Figure~\ref{FCE_results}a shows the numerical results obtained using the FCE approach based on the set of pre-trained networks. It is worth recalling that the network were not re-trained for the specific system configuration but simply extracted from LE. As in previous cases, the results were compared with the predictions from a finite element solution considered as ground truth. In addition, the prediction using a single BRNN network (hence without using FCE) was obtained. The direct comparison of the three results shows that the single network fail to accurately predict the bar response, while the FCE method provides a result within a 2\% margin of error compared to the FE model solution.

As previously mentioned, while FCE is a fundamental methodology to create complex physical models by interconnecting pre-trained networks, it is also a key approach to the efficient simulation of large scale systems. We illustrate this capability in the following example where the structure in case 3 was solved with different number of elements by using both the conventional FE method and FENA with FCE. The resulting performance is compared based on simulation (CPU) time. More specifically, Figure~\ref{FCE_results}b reports the results of the computational time needed to simulate the same physical system discretized with an increasing number of elements ($n$) ranging from 300 to 35700 in increments of 300. Each problem was solved using both an in-house FE model code and FENA. The decision to use an in-house FE code was motivated by the need to use the sample platform (i.e. Python) in order to enforce consistency of test conditions. Simulations were performed on the same machine, an Intel Xeon E5-2660 CPU with 256 Gb of memory. Figure~\ref{FCE_results}b presents the results in terms of the ratio $t_r=t_{\text{FENA}}/t_{\text{FE}}$ of the computational time of both FENA and the FE method. It appears that, under the specified conditions, FCE tends to be less efficient than the FE method for small size problems ($\approx \leq 10^4$), but it outperforms the FE method for large size problems.

Note that an important reason for the lower performance of FENA for small size systems is the time required to solve the minimization problem involved in FCE. The strategy used to achieve the continuity of different sections was not optimized for performance but simply to show the feasibility of the concept. It is easy to envision that, for smaller size systems, it would be a much more efficient approach to use larger size BRNN elements. Clearly, these are all considerations related to the formation of a comprehensive LE module capable of approaching a variety of problems. Also, other strategies could be envisioned to perform the concatenation that do not rely on a numerical optimization method. While the present paper focuses on introducing the concept of a network-based computational framework, future studies will have to focus more directly on performance improvements.

\subsection{Case 4: assessing accuracy and reliability via the MA}
\begin{figure}[!ht]
	\centering
	\includegraphics[width=.7\linewidth]{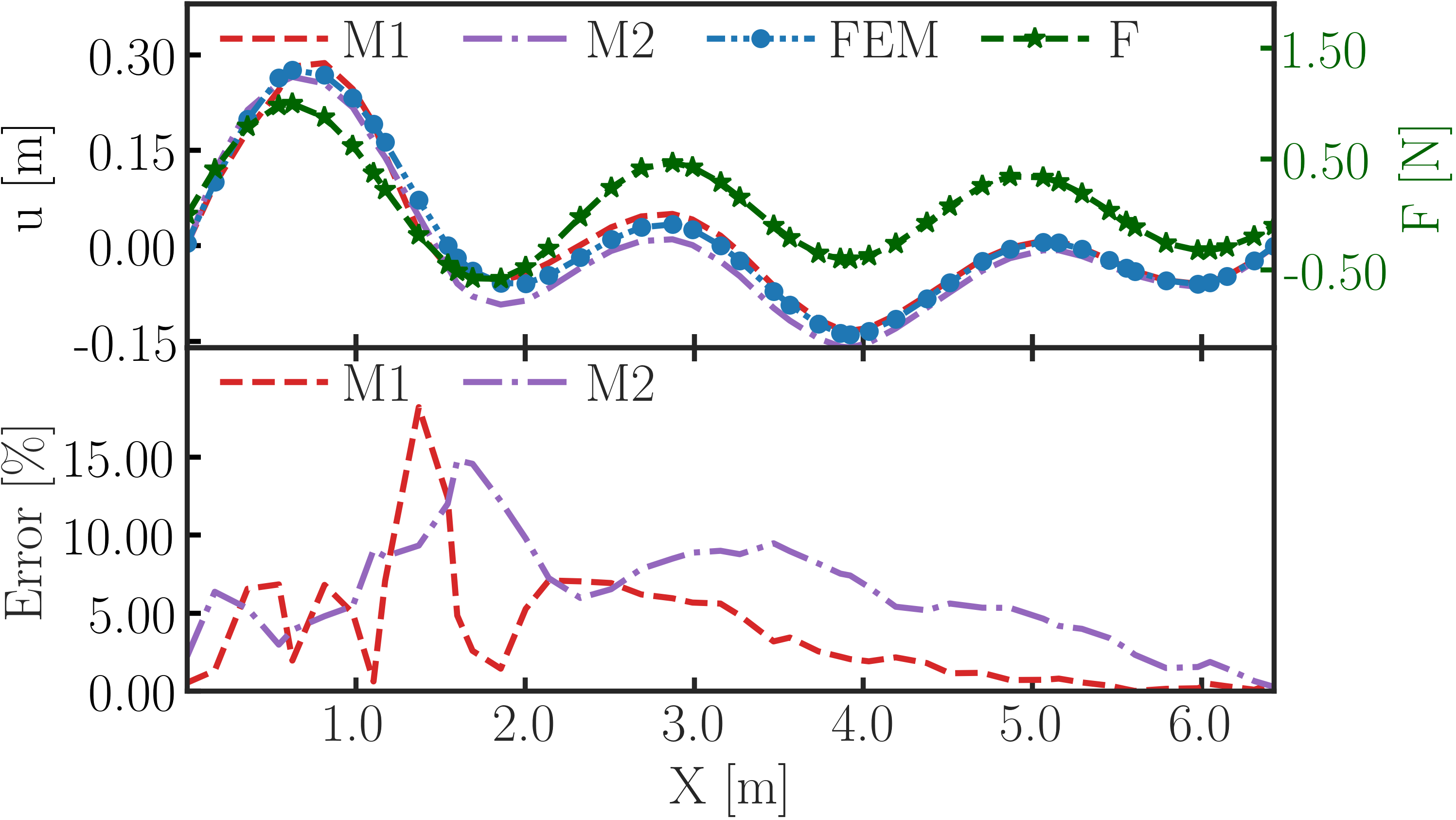}
	\caption{Numerical results showing the performance of models M1 and M2 for case 4 under the original (i.e. unscaled) input load. (top) The displacement profile is reported for both models M1 and M2 and for a reference solution obtained via a converged FE model (FEM). (bottom) The percentage relative nodal error obtained by comparing the predictions from models M1 and M2 with the reference FE solution. At this scale, some nodes exhibit an error larger than 10\%.}\label{Poor_pred}
\end{figure}

While previous sections illustrated the use of FENA and its performance, this section will show how the quality of the solution can be assessed via the MA module. This functionality is illustrated via case 4 that considers a bar subject to a distributed load discretized in forty-six nodal loads, as shown in Figure~\ref{Poor_pred}. This case can be solved using a single BRNN, hence applying a solution procedure equivalent to that discussed for cases 1 and 2. Figure~\ref{Poor_pred} shows the results in terms of calculated nodal displacements obtained from both models 1 and 2, and compared directly with the FE model solution. At some nodes, the percentage relative error is greater than $10\%$.\\
\textbf{Model assessment}: for case 4, the ESI index for the two models is $ESI_{M1}=0.62, ESI_{M2} = 0.27$. Compared to cases 1 and 2, there is a significant increase in ESI value, which can be a sign of inaccurate predictions. As discussed in \S~\ref{MA-description-BAR}, in these situations a possible strategy is to simulate the problem via multiple models at different scales and then to perform a voting operation in order to obtain an accurate prediction.

Considering the linearity characteristic of the problem, we can apply the combined linearity and similarity index. Thus, we solved the problem using the four models with input load scaled in two different ways: 1) the original scale, and 2) a reduced scale by a factor of 5. The resulting PPMCC matrices were calculated as: 
\begin{equation}\label{PPMCC}
	\begin{aligned}
		&\text{Scale}=\text{1.0:}&\begin{blockarray}{ccccc}
			\text{M1}&\text{M2}&\text{M3}&\text{M4}& \\
			\begin{block}{[cccc]c}\bigstrut[t]
				1&0.986&0.987&0.993&\text{M1} \\
				& 1    &0.983&0.977&\text{M2} \\
				&     &1   &0.984&\text{M3} \\
				\multicolumn{2}{c}{\text{\smash{\raisebox{1.5ex}{sym.}}}}&
				& 1.  &\text{M4} \bigstrut[b]\\
			\end{block}
		\end{blockarray}\ 	
		&\text{Scale}=\text{0.2:}&\begin{blockarray}{ccccc}
			\text{M1}&\text{M2}&\text{M3}&\text{M4}&\\
			\begin{block}{[cccc]c}\bigstrut[t]
				1&0.999&0.999&0.998&\text{M1} \\
				&1    &0.998&0.998&\text{M2} \\
				&     &1    &0.998&\text{M3} \\
				\multicolumn{2}{c}{\text{\smash{\raisebox{1.5ex} {sym.}}}}&     & 1   & \text{M4} \bigstrut[b] \\
			\end{block}
		\end{blockarray} \\	
	\end{aligned}
\end{equation}
similar to Eq.~\ref{PPMCC-P12}, each element of the matrices represents the correlation between model pairs.
By comparing the results of the two indices, we observe that while the correlation is relatively low when the input load at the original scale (i.e. scale=1) is used, the correlation increases significantly for the scaled down input (i.e. scale=0.2). Hence, we conclude that at scale=1 all the models are unreliable, while for the scaled input the results are reliable and their average should be considered as the model ensemble prediction. For case 4, the averaged scaled response yields a maximum nodal displacement error of 2\% (see \ref{sec: Case4 Suppresults} for a summary of the displacement and error values of each model at the two scales). Note that, even if only one of the four models was available, the linearity index could still be applied to different input scales in order to assess the accuracy and convergence of the model. 

\section{Conclusions}
This paper introduces the concept of network-based computational framework for the simulation of physical systems. The most remarkable aspect of this framework includes the ability to simulate systems made of multiple components by simply interconnecting pre-trained networks available in a library and acting as surrogate models for specific system's functionalities. In other terms, pre-trained networks can be connected to assemble a model of a system without requiring any further training after the assembly phase. The close conceptual analogy between this property and the strategy used in classical finite element analysis to build complex models, lead to the definition of Finite Element Network Analysis (FENA). This terminology highlights the modular nature of the approach (from this perspective, analogous to FEA) where traditional finite elements are replaced by pre-trained network elements.

Another outstanding property of FENA is the potential for extreme computational efficiency that draws directly from the properties of trained neural networks. This efficiency can be particularly significant when approaching traditionally challenging problems including multiscale and large size (i.e. high number of degrees of freedom) models. These unique properties of FENA are possible thanks to a library of fundamental building blocks (i.e. the finite element networks) built on the architecture of deep bidirectional recurrent neural networks (BRNN). BRNNs allow important features, like variable input sequence size and bidirectional flow of information, that serve as the foundation to enable physical modeling. As an example, forcing and boundary conditions are problem-dependent characteristics that cannot be generalized during the training phase and that, with classical network architectures, would require dedicated and repeating training. 

This paper focuses on presenting the fundamental concept behind FENA and its core structure. While the framework is general and can be applied potentially to the simulation of any physical system (given the development of the appropriate library of surrogate elements to simulate the physics of interest), we exemplified its application for the static analysis of 1D elastic structure under axial loads.

The concept was tested via numerical examples designed to showcase unique functionalities of the framework ranging from the application of different loading and boundary conditions without the need for network re-training, to the ability to concatenate finite networks together in order to assemble models of interconnected systems. The direct comparison of the numerical results with those obtained from a traditional finite element approach indicated the remarkable capability of FENA to accurately solve generalized problems under the most diverse system sizes, external loads, and boundary conditions. While the present framework was developed and validated for the specific case of 1D structural analysis, it is possible to envision that FENA could be generalized to simulate more complex and higher dimensional systems of interest in multiple branches of computational physics. Of course, the practical application and performance of FENA within these different contexts will have to be confirmed via focused studies.

\section*{Acknowledgments}
The authors gratefully acknowledge the financial support of the National Science Foundation (NSF) under grant CAREER \#1621909.
\section*{Author contributions}
M. J. developed the theoretical and numerical models and performed the numerical study. F. S. conceived the research and supported its development. Both authors contributed equally to writing the manuscript.

\section*{Competing interests }
The authors declare no competing interest.

\bibliographystyle{elsarticle-num} 
\bibliography{Refs}

\newpage

\appendix
\setcounter{equation}{0}
\setcounter{figure}{0}
\setcounter{table}{0}

\section{Recurrent Neural Networks}\label{RNN_LSTM_dis}
This section briefly describes the main features and operating principles of recurrent neural networks (RNN) and long-short term memory (LSTM) cells.  

\begin{figure}[!htb]
	\centering
	\includegraphics[width=.5\linewidth]{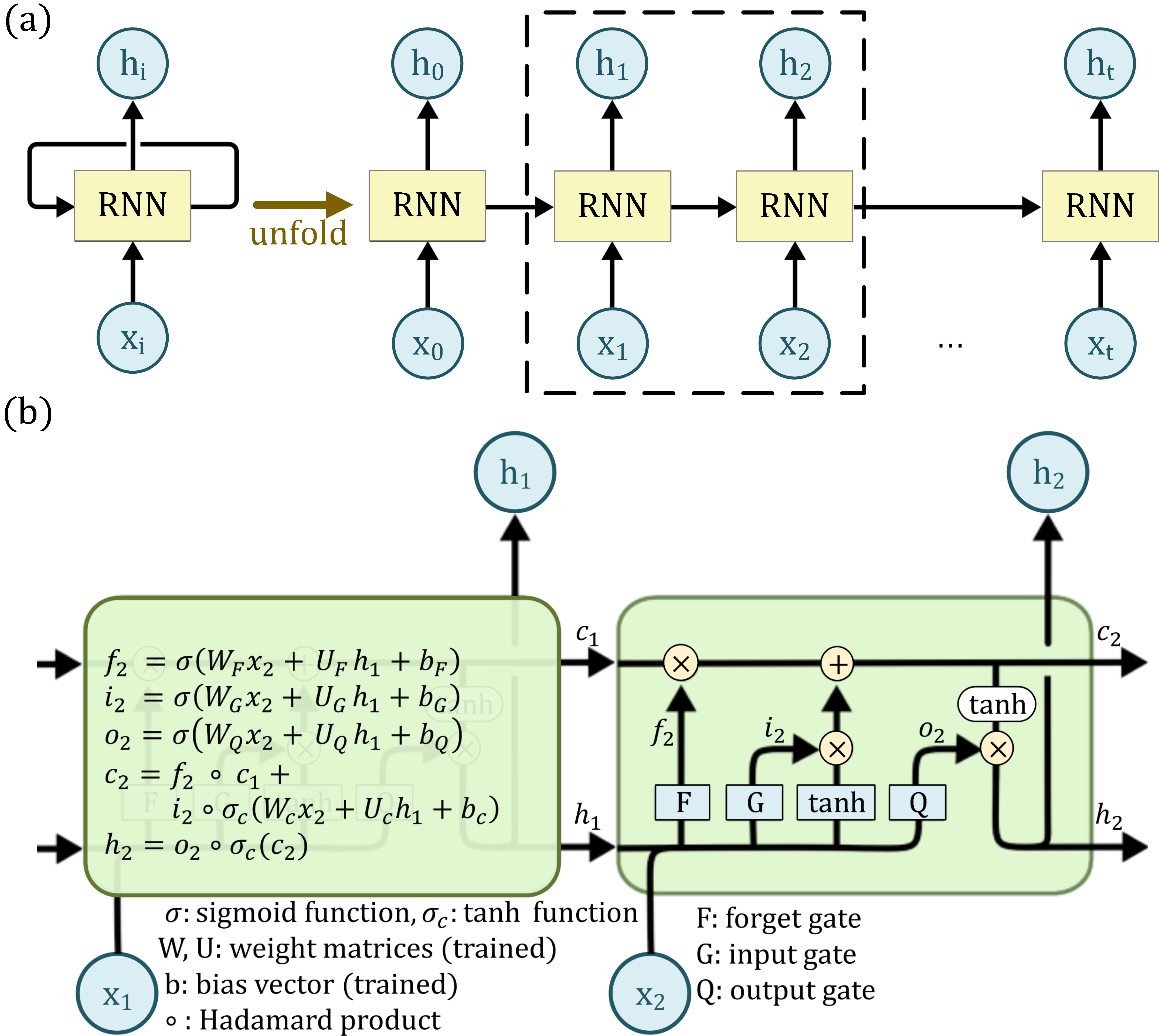}
	\caption{
		(a) View of the folded and unfolded architecture of a RNN. The RNN cell output ($h_i, \ i = 0,1,2, ..., t$) at state $i$  is calculated using the input $x_i$ and feedback from the previous state ($h_{t-1}$). The feedback is updated to be used at the state $i+1$. A more detailed view of the RNN blocks (dashed area) is presented in part (b) where LSTM cells are used as RNN blocks. (b) Detailed internal structure of a LSTM cell. The left block shows the equations relating input and output of the LSTM cell at state 2. $x_2$, previous step output $h_{1}$ , and cell state $c_{1}$ contributions to the output $h_{2}$ are controlled by input $G$, output $Q$, and forgot $F$ gates.}\label{RNN_schem} 
\end{figure}

Figure~\ref{RNN_schem}a shows the architecture of a RNN both in a folded (left) and an unfolded (right) representation. The latter highlights more clearly how RNN maintains a memory of the previous states as the information is processed. The block \textit{RNN} shows the main core of the network which receives input $x_i$ and output at state $i-1$ to calculate the output $h_i$ at the $i$-${th}$ step of the input sequence $x_i$. 

Although this basic form of RNN~\cite{RNN-Simple1990} only considers the output at the previous step (hence it exhibits a short memory), longer sequences (i.e. longer memory) can be achieved by replacing the RNN blocks in Figure~\ref{RNN_schem}a with Long-Short Term Memory (LSTM) cells~\cite{Gers1999}. Besides the ability to receive feedback from previous output, LSTM has an additional internal parameter called cell state $c$ to carry information from the previous states. The flow of information between states is regulated by three internal gates: 1) input gate $G$, 2) output gate $Q$, and 3) forget gate $F$, as shown in Figure~\ref{RNN_schem}b. $G$, $Q$, and $F$ use a sigmoid activation function, that takes only values between zero and one, to control the amount of information that is allowed to flow (a gate value of zero means no  information is passed while a gate value of one allows the exact value of information to flow). As an example, it is seen in the right cell in Figure~\ref{RNN_schem}b that $c_1$ is first regulated with the forget gate output, $f_2$, before being used in
state 2. Gate controlled flow of information in LSTM results in a strong ability to learn long term sequence dependencies. Figure~\ref{RNN_schem}b presents the internal structure of a typical LSTM and the input-output mathematical relationship behind it. A detailed description of the LSTM cell can be found in~\cite{Goodfellow-et-al-2016}.

\setcounter{equation}{0}
\setcounter{figure}{0}
\setcounter{table}{0}

\section{Network architecture design and training}\label{netstructure}
\subsection{Network architecture}
The network architecture used to build the fundamental surrogate physical models in LE contains both fully connected and BRNN layers. In the following we describe the network architecture used in building LE for elastic bar elements.

As discussed in the main text, LE members simulate the response of a physical system by discretizing the computational domain into elements, building a sequence of data from each element properties, and predicting the corresponding output sequence. In our specific example where the framework is used to simulate the static response of 1D structures under an external axial load, the total number of elements and their lengths are determined by: 1) the location of the applied external load (following the general rationale that a node is needed at any point of application of the load), 2) the physical length of the 1D bar, and 3) the number of elements that the network model is trained to simulate. In the axial bar problem, the cross-sectional area $A_i$, the Young's modulus $E_i$, the start and end point coordinates of each element (which determines the element length $l_i$), and the corresponding applied loads ($F_{i,l}$ and $F_{i,r}$) form the input sequence to the network. The left and right boundary conditions are given to the network as additional input that determine the initial hidden states (HS) of the LSTM cells in each direction. Theses physical parameters can span different orders of magnitude thus, in order to eliminate scaling issues, the data are normalized with a nonlinear function given by: 
\begin{equation}\label{normalization}
	N(x, x_{max}, x_{min}) = \frac{log(x) - log(x_{min})}{log(x_{max}) - log(x_{min})} 
\end{equation}
where $x_{max}$ and $x_{min}$ are the maximum and minimum values of the input parameter $x$. 

The normalized input sequence ($E_i$, $A_i$, $l_i$, $F_{i,{l}}$, $ F_{i,{r}} $; $i= 1,2,...,n$) is then passed through three fully connected layers with size [14,14,42]. The first layer uses $tanh(x)$ as activation function while the remaining two use the E-swish activation function~\cite{E-swish-alcaide2018} defined as:

\begin{equation}\label{activation}
	f(x) = \frac{-\beta x}{1+e^{-x}}, \ \beta = 1.5  
\end{equation}

The E-swish function has a linear behavior for large positive input values which may cause numerical instability (explosion of the cost function gradient) when it is used in recurrent layers. On the other hand, $tanh$ function saturates over a large section of its domain which adversely affects prediction accuracy. We used a combination of both activation functions in fully connected layers in order to prevent both the numerical instability and the saturation of layers. 

\begin{figure}[!htbp]
	\centering
	\includegraphics[width=.5\linewidth]{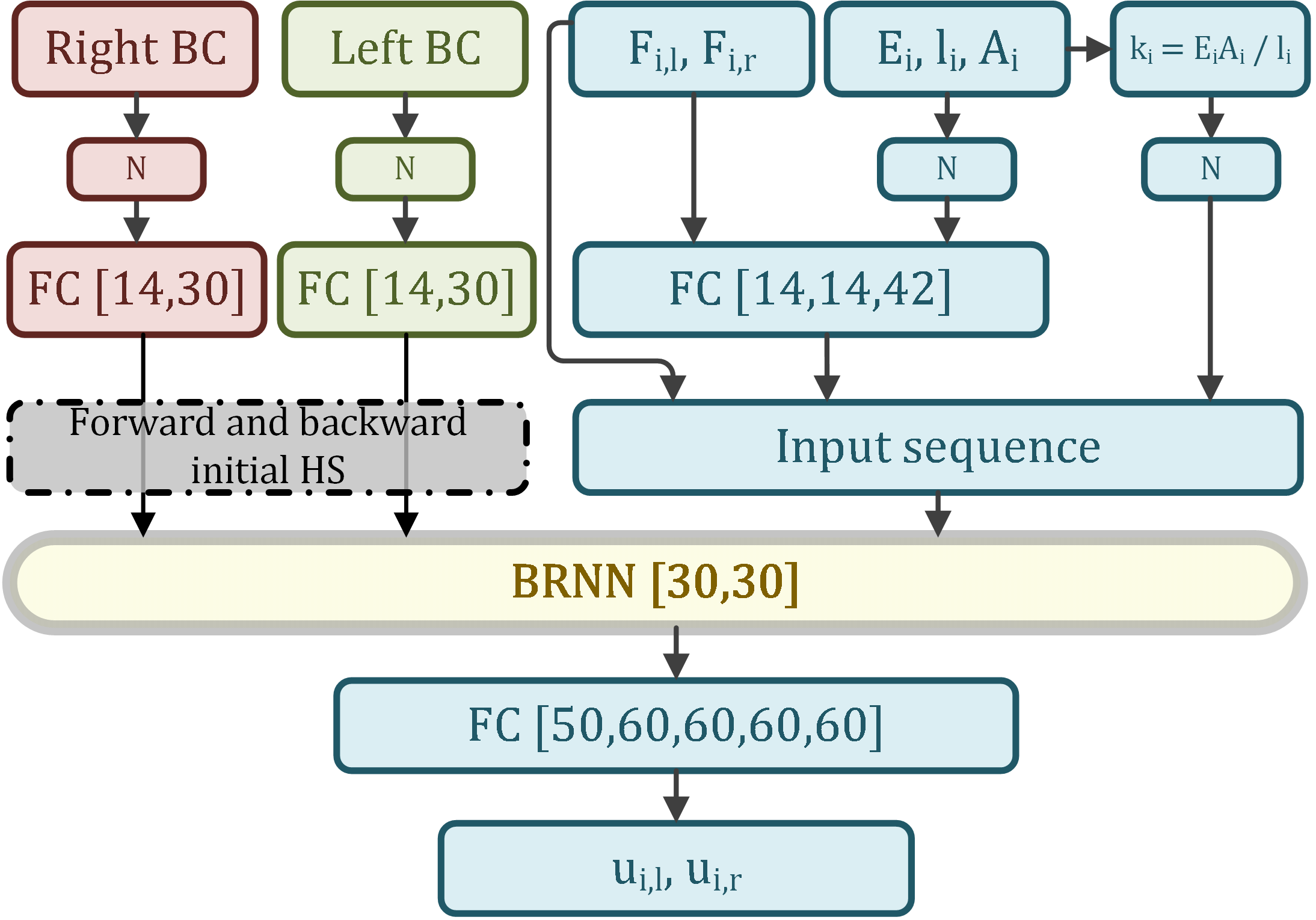}
	\caption{Network architecture to create the surrogate models for the static response of elastic bars. The network receives a sequence of properties ($F_{i,l}$, $F_{i,r}$, $A_i$, $E_i$, and $l_i$; $i=1,2,...,n$) and boundary conditions (left and right BC) as input. The input is first normalized by the normalization function $N$ and then passed to fully connected layers(FC). The numbers in brackets indicate the size of each layer. The network has one BRNN layer with 30 LSTM cells in each direction.} 
	\label{Net-Structure}
\end{figure}

The input sequence is processed by the BRNN layer, which has 30 LSTM cells in each forward and backward direction. The LSTM cells have the role of learning the sequential logic, and relating input sequence to the unknown nodal displacement output. As previously discussed, the initial states are strictly connected to the boundary conditions of the problem. In our specific problem, the boundary conditions are defined via a discrete local stiffness at the boundary nodes. The (left and right) endpoint stiffness values are used to calculate the initial value of the HS for both the forward and backward LSTM cells inside the BRNN layer. Initializing the HS of the 30 LSTM cells directly with the same value (boundary stiffness) limits the capability of these cells to learn the relation between input and output sequences, hence limiting the prediction accuracy. Thus, the left and right boundary stiffness values are first passed to two consecutive fully connected layers (with trainable weights) of size [14,30] that are used to identify 30 different initial hidden states for the LSTM cells in each direction. The output of the bidirectional layer is then passed through five fully connected layers with size [50,60,60,60,60] and E-swish activation function. The network output layer size is equal to 2 and has a linear activation function. A schematic of the network architecture is provided in Figure~\ref{Net-Structure}. Note that the network layers were determined via a trial and error procedure that placed particular attention on the numerical stability of the training process and on the overall prediction accuracy.

It is worth mentioning that the case of displacement boundary conditions can be treated in an equivalent way to the case of stiffness boundary conditions. In other terms, the boundary displacements can be provided as input to the network in order to initialize the hidden states of the forward and backward recurrent cells. The network is then trained using sample problems under assigned boundary displacements. At this stage, we can also speculate that a similar approach could also be followed in the case of a mixed boundary condition (i.e. boundaries that can be specified either as stiffness or displacement boundaries). These network elements, trained for different boundary conditions, can be stored in the LE and can be selected based on the specific problem conditions. The procedures for data set generation (except changing the type of boundary conditions), network training, and concatenation remain the same regardless of the specific boundary condition being selected.

\subsection{Multi-objective training of the network}\label{Training}\leavevmode\\
\textbf{\underline{Loss function:}}
The most commonly used loss function in regression problems solved via neural networks is the mean square error (MSE)~\cite{Goodfellow-et-al-2016}. The MSE loss function tends to be biased towards elements with higher magnitude and loses accuracy when the magnitude decreases. Thus, training the network by using exclusively the MSE tends to reduce the accuracy of the prediction for certain output ranges. In order to overcome these issues and obtain accurate predictions, a tailored loss function made up of three different terms defined in Eq.~\ref{loss_function} is used during training.\\
\begin{equation}\label{loss_function}
	\begin{split}
		Loss = & \underbrace{\frac{\mathbb{\omega}_1}{\text{NF}} \times \frac{1}{n} \sum_{i=1}^n \left(u^{pred}_{i,\{l,r\}}-u^{true}_{i,\{l,r\}}\right)^2}_{\text{normalized MSE}} + 
		\underbrace{ \frac{\mathbb{\omega}_2}{\text{NF}} \times \frac{1}{n}\sum_{i=1}^n {
				\left(u^{pred}_{i,{r}}-u^{true}_{i+1,l}\right)^2}
		}_{\text{continuity}} + \\
		& \underbrace{ \frac{\mathbb{\omega}_3}{\text{NF}} \times\frac{1}{n} \sum_{i=1}^{n} { \left( \frac{ \left( \left({u^{pred}_{i,r} + u^{pred}_{{i+1},l}} \right) / 2 \right) \times \left(k_i + k_{i+1} \right) - k_i u^{pred}_{i,l} -k_{i + 1 } u^{pred}_{{i+1},{r}} - F_{i,{l}} }{ {\left(k_{i} + k_{i+1}\right)}/{2} } \right) ^2}}_{\text{normalized physical law}} 
	\end{split}
\end{equation}
\noindent where $\text{NF} = \frac{1}{n}\sum_{i=1}^n\sum_{j=l,r} \left(\left|{u^{true}_{i,j}}\right|+\epsilon \right)^2$,
$u^{pred}_\square$ and $ u^{true}_\square$ are the network predicted and exact displacement values, $k_i$ is the stiffness of the $i$-${th}$ element, $\mathbb{\omega}_i$ is a weight factor, and $\epsilon = 1^{-8}$ is a regularization factor. The first term, labeled \textit{normalized MSE}, is the normalized prediction mean square error. The terms labeled \textit{continuity} and \textit{normalized physical law} enforce physical constraints based on both the  intrinsic nature and response of the system. More specifically, the normalized physical law term implements the equilibrium equation for the 1D structure, while the continuity term enforces the continuity of displacements at the nodes of adjacent elements.
We highlight that the normalization factor (NF) plays a key role in the network training. This factor scales each sample problem in a training batch based on the average value of the nodal displacements, hence removing the scale inconsistency between different sample problems in the batch. 
In the absence of normalization, the process of  the network weights update in each training step tend to be biased towards those data samples associated with higher displacement values since the cost function and its gradient are mostly affected by those data samples. Hence, the prediction accuracy would tend to decrease for domains with lower amplitude of the physical response. \\
\noindent \textbf{\underline{Data set generation and network training:}}
The data set used in this paper is generated by an in-house finite element code. We selected a set of input parameters ranges reported in Table~\ref{table:c1} and used them to generate the training data set. We assumed that each sample problem had a uniform area $A$ and Young's modulus $E$, which were randomly selected from a uniform distribution within the ranges presented in Table~\ref{table:c1}. Note that this random strategy allows obtaining network that are more general and not biased towards a specific range of the data set. The total number of elements $n$ and length of each element $l_i$ in a data sample are selected from a uniform distribution within the corresponding ranges. Note that the intrinsic linearity of the problem at hand (recall we are addressing the linear elastic response of a 1D structural element), allows using a simple and intuitive scaling process of the range of parameters and of the output of the trained network in order to simulate problems outside the range indicated in Table~\ref{table:c1}. In other terms, when the problem is linear, the ranges used for training are not very restrictive for the performance of the network because a given problem can be solved with scaled input or properties and the simulated response can always be rescaled to the range of interest. The analysis of case 4 in the main text, addressed this exact condition.

\begin{table}[!bhtp]
	\centering
	\resizebox{.45\columnwidth}{!}{
		\begin{tabular}{lr} 
			Input & Range  \\ 
			\hline
			E [$kPa$] & [$50, 200$] \\ 
			A [$m^2$] & [$\medmuskip=0mu 5\times 10^{-5}, 2\times 10^{-4}$]  \\ 
			l [$m$]& [$\medmuskip=0mu 5 \times 10^{-2} , 2 \times 10^{-1}$]   \\ 
			F [$N$] & [$-1, 1$]  \\ 
			Boundary Stiffness [$N/m$] & [$\medmuskip=0mu 1.25\times 10^{-1}, 1.4\times 10^4$] \\ 
			n & [$10, 60$]  \\ 
			\hline
			
	\end{tabular}}
	\caption{Parameters used to generate the training data set.
	}
	\label{table:c1}
\end{table}

The batch size was set to 95 for the network training. Thirty batches of data (sample problems) were generated for each total number of elements $n\in [10,60]$ hence resulting in $30\times51\times95 = 145350$ total training sample problems in the data set. The data set size and batch size were chosen via a trial and error process considering the network accuracy. Two training data set, set \#1 and \#2, were generated. The first data set was used to train model M1 and M2 and the second one was to train model M3 and M4. Individual randomly-generated data sets were used to guarantee statistical independence of the models M1-2 and M3-4. A validation data set, 15\% the size of the training set, was generated under the same conditions of the training set.  

The models were built using python \textit{Keras} and \textit{Tensorflow} packages. The training process involving different training data sets and different sequence sizes leveraged the \textit{data generator} function and the Keras \textit{fit\_generator} training method. The data generator loads batches of sample problems (having the same sequence length from the training data set) that are used by the \textit{fit\_generator} at each training iteration. One training epoch is completed once all the samples with different sequence sizes are loaded (batch by batch) and used by the \textit{fit\_generator}. The order of sequence sizes of the data batches is shuffled at the beginning of each epoch. The models were trained using the ADAM\cite{adam} optimization algorithm. A total of 360 epochs were used together with an initial learning rate of $.0001$. The learning rate was divided by a factor of two every 120 epochs. The cost function is defined in Eq.~\ref{loss_function}.

\setcounter{equation}{0}
\setcounter{figure}{0}
\setcounter{table}{0}

\section{Supplementary Results}\label{sec:supplementary-results}
\subsection{Training of models M1-4}
As described in the main manuscript, four models (labeled M1, M2, M3, and M4) were built and trained to model the response of a 1D bar under axial load. The training conditions for these models are listed in the Table~\ref{table:model_description_apndx}.  

\begin{table}[!h]
	\centering
	\resizebox{.45\columnwidth}{!}{
		\begin{tabular}{lrr} 
			Model &Training data& Loss function\\
			name &set No.& weights\\ 
			\hline
			M1& \#1&  $\medmuskip=0mu \omega_1=\omega_2=1, \omega_3 = 0$\\ 
			M2& \#1&  $\medmuskip=0mu \omega_1=\omega_2=\omega_3 = 1$\\ 
			M3& \#2&  $\medmuskip=0mu \omega_1=\omega_2=1, \omega_3 = 0$\\ 
			M4& \#2&  $\medmuskip=0mu \omega_1=\omega_2= \omega_3 =1$\\ 
			\hline
		\end{tabular}
	}
	\caption{Data set and weights of the loss function used to train the different models in LE.}
	\label{table:model_description_apndx}
\end{table}

\begin{figure}[!h]
	\centering
	\includegraphics[width=.6\linewidth]{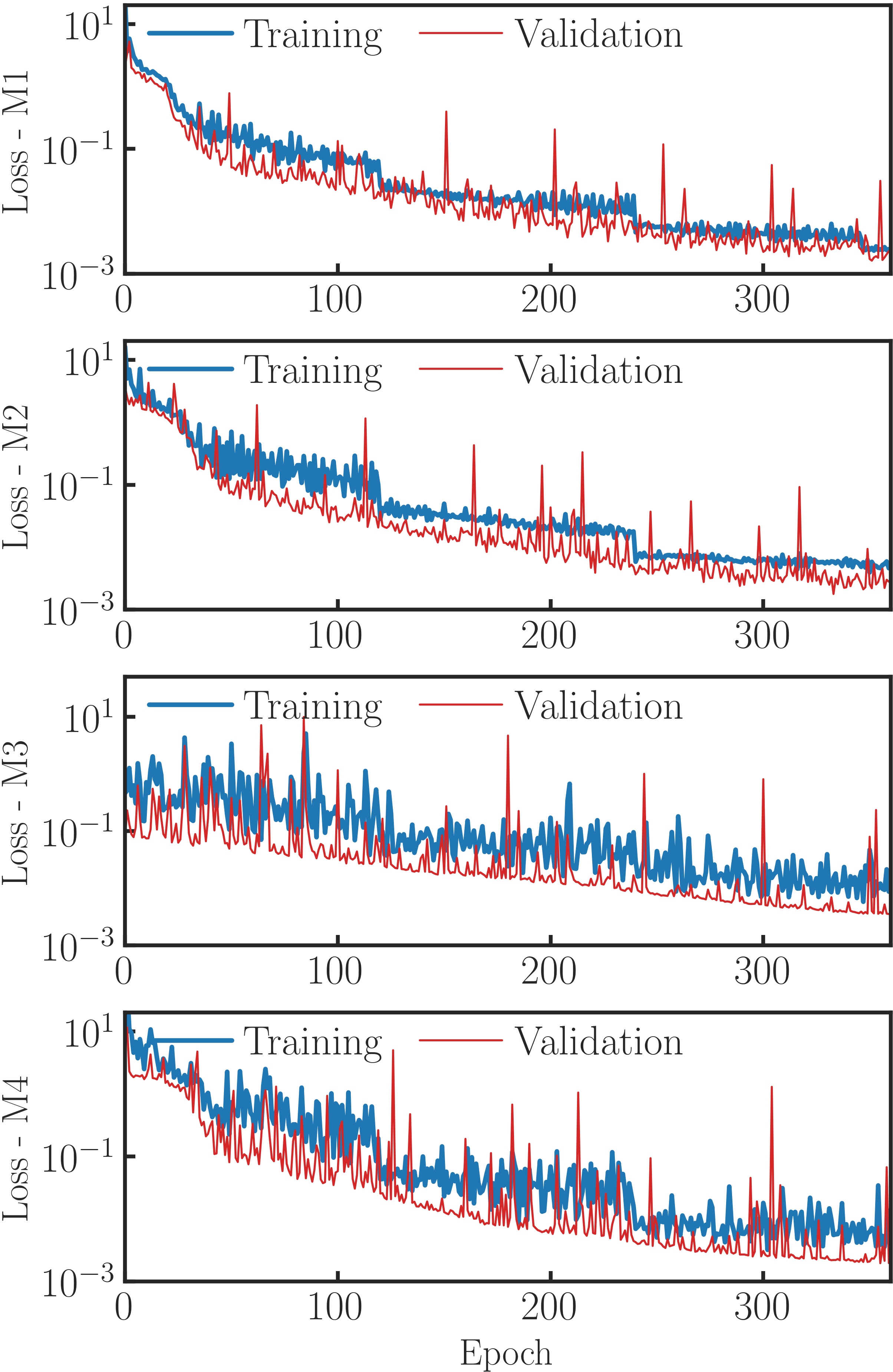}
	\caption{Loss function versus the epoch number for both the training and validation data sets. Results are reported for the four models (M1-4) used in the LE.} \label{Loss-Train}
\end{figure}

All the models were trained using the same training hyper-parameters described in \S~\ref{Training}. Figure~\ref{Loss-Train} shows the trend of the loss function versus the epoch number during the training phase. The fact that the loss function for both the training and validation data sets converges to similar values indicates that all the trained models are generalized and not over-fitted on the training data set.

\subsection{Case 1 and 2: static response under distributed loads - results for M3 and M4 models}\label{sec: M3-4Suppresults}
\begin{figure}[!h]
	\centering
	\includegraphics[width=.7\columnwidth]{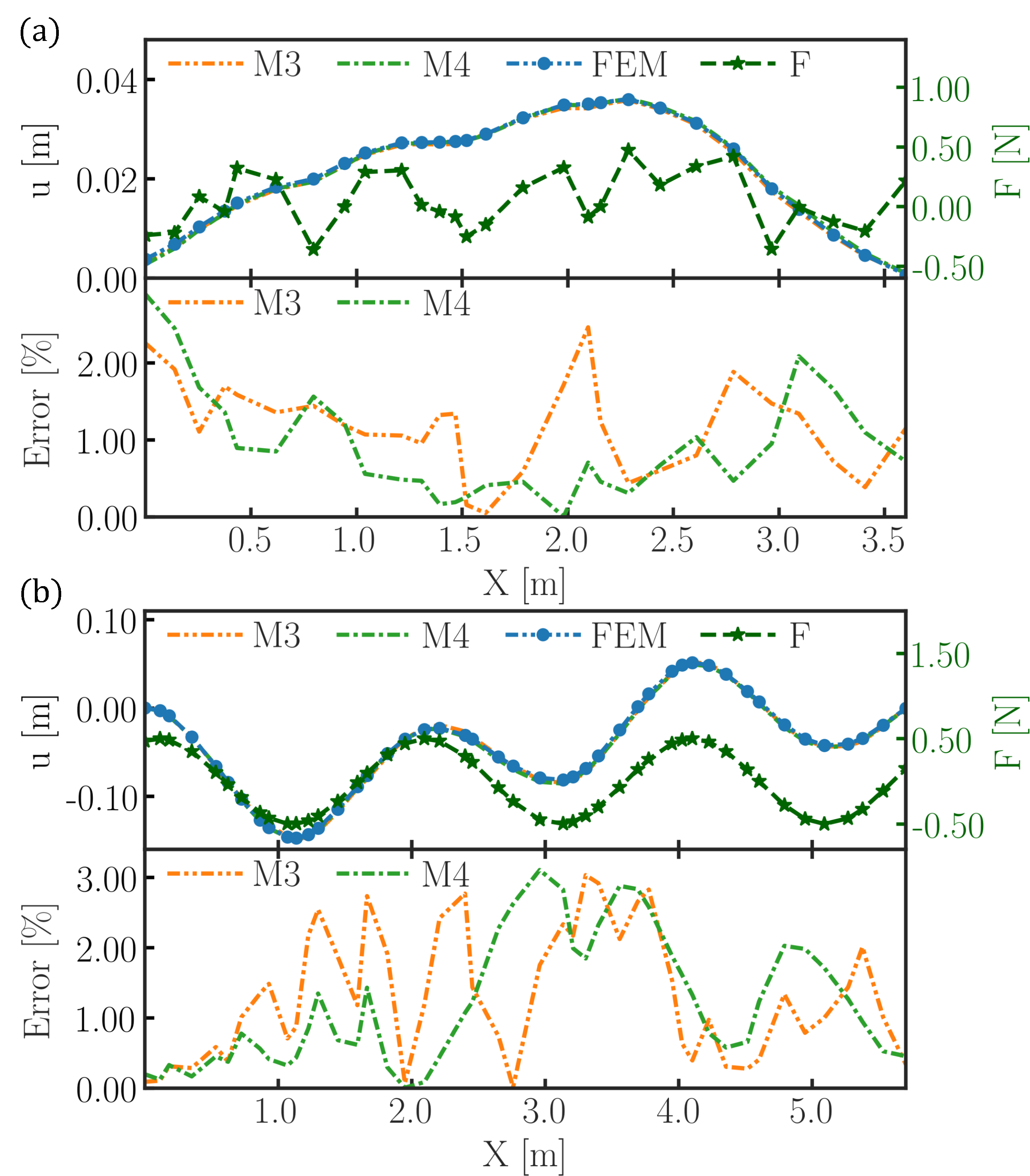}
	\caption{(a) Numerical results of cases 1 solved using models M3 and M4. (top) Nodal displacement profile compared with a converged FE model (FEM) under the applied load $F$. (bottom) Nodal percentage relative error. (b) Numerical results of cases 2 solved using models M3 and M4. (top) Nodal displacement profile compared with a converged FE solutions under the applied load $F$. (bottom) Nodal percentage relative error. Note that, in both plots, star markers indicate the nodes location.}\label{R1_M34}
\end{figure}

This section complements the results presented in the main manuscript by reporting the numerical predictions for cases 1 and 2 obtained with model M3 and M4. Other than using different models, the remaining conditions were equivalent to the tests performed with model M1 and M2. The results are shown in terms of displacement profile and percentage relative error in Figure~\ref{R1_M34}a-b. The predictions from both models were accurate within a 3\% error. In case 1, the ESI index is equal to $ESI_{M3}=0.110$ and $ESI_{M4}=0.088$ for model M3 and M4, respectively. 
In case 2, the ESI index is equal to $ESI_{M3}=0.059$ and $ESI_{M4}=0.017$. 

\subsection{Supplementary results for Case 4}\label{sec: Case4 Suppresults}

\begin{figure}[!t]
	\centering
	\includegraphics[width=.7\linewidth]{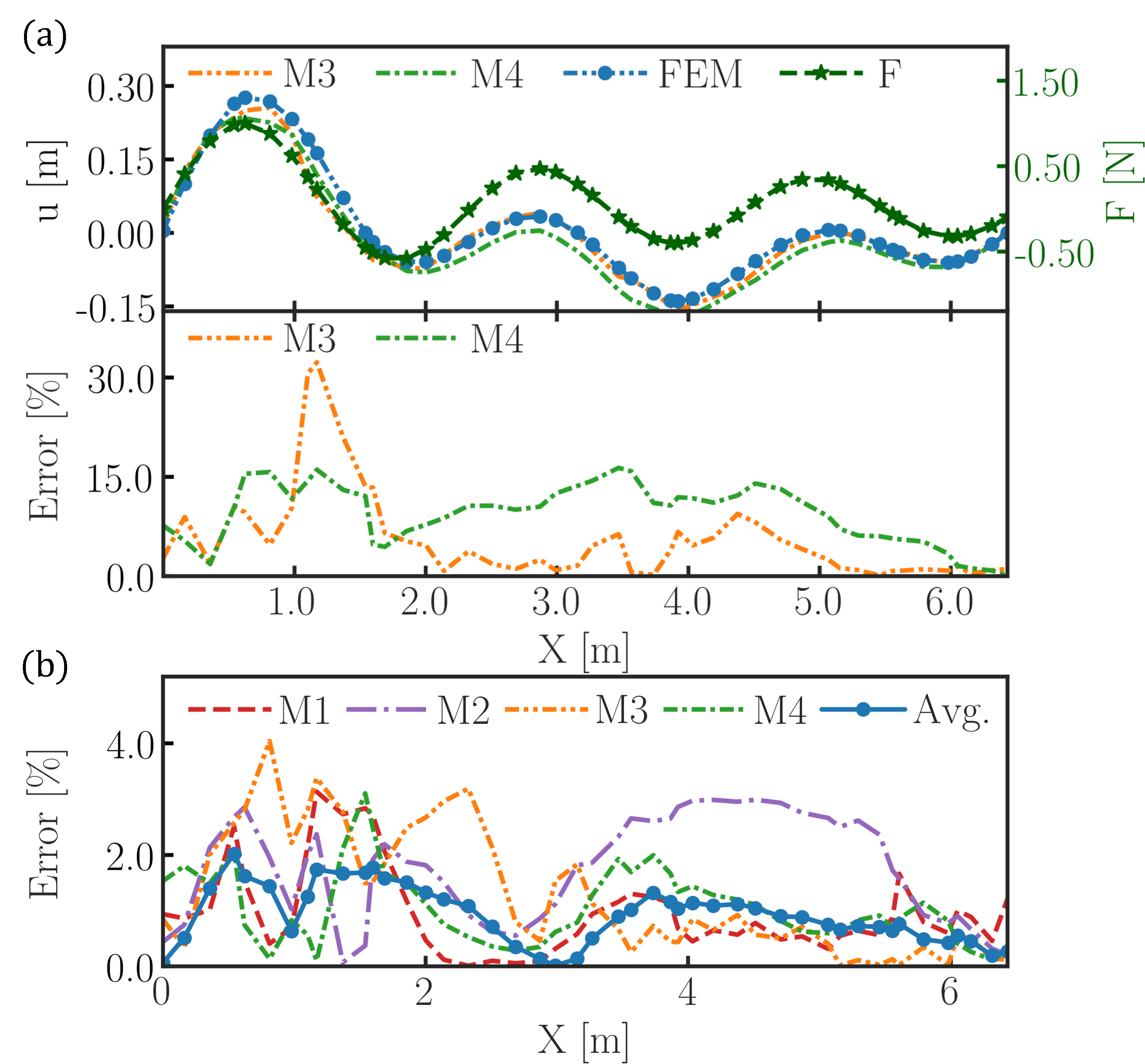}
	\caption{(a) Numerical predictions from models M3 and M4 for case 4 under unscaled input load $F$. The maximum prediction errors exceeds 15\% in both models. The solution from the converged FE model is also reported. (b) Nodal percentage relative error for the four models and for the averaged response from the scaled input load.}
	\label{Poor_pred_M34}
\end{figure}

Similarly, models M3 and M4 were used also to obtain numerical predictions for case 4. In this case, the ESI index was equal to $ESI_{M3}=1.714$ and $ESI_{M4}=0.78$. This index shows a significant increase compared to the ESI values in the cases 1 and 2 hence suggesting that the predicted responses are not accurate and reliable. Figure~\ref{Poor_pred_M34}a shows the results in terms of nodal displacement profile obtained from models M3 and M4, and from the FE model solution. At certain nodes, the relative error is greater than $15\%$. Figure~\ref{Poor_pred_M34}b shows the nodal percentage relative error of all four models for the scaled input load. All the models predict the response with less than a 4\% prediction error. Low prediction error values were expected due to the highly correlated predictions for the scaled applied load. For the scaled input, the results of all models are reliable and their average should be considered as the model ensemble prediction. The nodal error of averaged response is also presented in Figure~\ref{Poor_pred_M34}b. It is seen that the averaged response nodal error is less than 2\% which has a significant decrease compared to the case of unscaled input. Note that the reported error values are normalized. Hence, scaling up the results to original values will not change the nodal error values. 

\end{document}